\documentstyle[twoside]{article}

\catcode`\@=11
\long\def\@makefntext#1{ 
\protect\noindent \hbox to 3.2pt {\hskip-.9pt
$^{{\eightrm\@thefnmark}}$\hfil}#1\hfill} 

\def\thefootnote{\fnsymbol{footnote}}
 \def\@makefnmark{\hbox to 0pt{$^{\@thefnmark}$\hss}}  

\def\ps@myheadings{\let\@mkboth\@gobbletwo
\def\@oddhead{\hbox{} 
\rightmark\hfil\eightrm\thepage}
\def\@oddfoot{}\def\@evenhead{\eightrm\thepage\hfil 
\leftmark\hbox{}}\def\@evenfoot{}
\def\sectionmark##1{}\def\subsectionmark##1{}}

\oddsidemargin=\evensidemargin
\renewcommand{\thefootnote}{\fnsymbol{footnote}}

\newcounter{sectionc}\newcounter{subsectionc}\newcounter{subsubsectionc}
\renewcommand{\section}[1] {\vspace{12pt}\addtocounter{sectionc}{1}
\setcounter{subsectionc}{0}\setcounter{subsubsectionc}{0}\noindent
	{\tenbf\thesectionc. #1}\par\vspace{5pt}}
\renewcommand{\subsection}[1] {\vspace{12pt}\addtocounter{subsectionc}{1}
	\setcounter{subsubsectionc}{0}\noindent
	{\bf\thesectionc.\thesubsectionc. {\kern1pt \bfit #1}}\par\vspace{5pt}}
\renewcommand{\subsubsection}[1] {\vspace{12pt}\addtocounter{subsubsectionc}{1}
	\noindent{\tenrm\thesectionc.\thesubsectionc.\thesubsubsectionc.
	{\kern1pt \tenit #1}}\par\vspace{5pt}}
\newcommand{\nonumsection}[1] {\vspace{12pt}\noindent{\tenbf #1}
	\par\vspace{5pt}}

\newcounter{appendixc}
\newcounter{subappendixc}[appendixc]
\newcounter{subsubappendixc}[subappendixc]
\renewcommand{\thesubappendixc}{\Alph{appendixc}.\arabic{subappendixc}}
\renewcommand{\thesubsubappendixc}
	{\Alph{appendixc}.\arabic{subappendixc}.\arabic{subsubappendixc}}

\renewcommand{\appendix}[1] {\vspace{12pt}
        \refstepcounter{appendixc}
        \setcounter{figure}{0}
        \setcounter{table}{0}
        \setcounter{lemma}{0}
        \setcounter{theorem}{0}
        \setcounter{corollary}{0}
        \setcounter{definition}{0}
        \setcounter{equation}{0}
        \renewcommand{\thefigure}{\Alph{appendixc}.\arabic{figure}}
        \renewcommand{\thetable}{\Alph{appendixc}.\arabic{table}}
        \renewcommand{\theappendixc}{\Alph{appendixc}}
        \renewcommand{\thelemma}{\Alph{appendixc}.\arabic{lemma}}
        \renewcommand{\thetheorem}{\Alph{appendixc}.\arabic{theorem}}
        \renewcommand{\thedefinition}{\Alph{appendixc}.\arabic{definition}}
        \renewcommand{\thecorollary}{\Alph{appendixc}.\arabic{corollary}}
        \renewcommand{\theequation}{\Alph{appendixc}.\arabic{equation}}
        \noindent{\tenbf Appendix \theappendixc #1}\par\vspace{5pt}}
\newcommand{\subappendix}[1] {\vspace{12pt}
        \refstepcounter{subappendixc}
        \noindent{\bf Appendix \thesubappendixc. {\kern1pt \bfit #1}}
	\par\vspace{5pt}}
\newcommand{\subsubappendix}[1] {\vspace{12pt}
        \refstepcounter{subsubappendixc}
        \noindent{\rm Appendix \thesubsubappendixc. {\kern1pt \tenit #1}}
	\par\vspace{5pt}}

\newcommand{\textlineskip}{\baselineskip=13pt}
\newcommand{\smalllineskip}{\baselineskip=10pt}

\def\eightcirc{
\begin{picture}(0,0)
\put(4.4,1.8){\circle{6.5}}
\end{picture}}
\def\eightcopyright{\eightcirc\kern2.7pt\hbox{\eightrm c}}

\newcommand{\copyrightheading}[1]
	{\vspace*{-2.5cm}\smalllineskip{\flushleft
	{\eightrm Modern Physics B, #1}\\
	{\eightrm $\eightcopyright$\, World Scientific Publishing
	 Company}\\
	 }}

\newcommand{\pub}[1]{{\begin{center}\eightrm\smalllineskip
	Submitted #1\\
	\end{center}
	}}

\def\abstracts#1#2#3{{
	\centering{\begin{minipage}{4.5in}\baselineskip=10pt\eightrm
	\centerline{ABSTRACT}
	\parindent=0pt #1\par
	\parindent=15pt #2\par
	\parindent=15pt #3
	\end{minipage} }\par}}

\renewenvironment{thebibliography}[1]			
	{\ninerm
	 \baselineskip=11pt				
	 \begin{list}{\arabic{enumi}.}
	{\usecounter{enumi}\setlength{\parsep}{0pt}
	 \setlength{\leftmargin 17pt}{\rightmargin 0pt}	
	 \setlength{\itemsep}{0pt} \settowidth		
	{\labelwidth}{#1.}\sloppy}}{\end{list}}

\newcounter{itemlistc}
\newcounter{romanlistc}
\newcounter{alphlistc}
\newcounter{arabiclistc}

\newcommand{\fcaption}[1]{
        \refstepcounter{figure}
        \setbox\@tempboxa = \hbox{\eightrm Fig.~\thefigure. #1}
        \ifdim \wd\@tempboxa > 5in
           {\begin{center}
        \parbox{5in}{\eightrm \smalllineskip Fig.~\thefigure. #1 }
            \end{center}}
        \else
             {\begin{center}
             {\eightrm Fig.~\thefigure. #1}
              \end{center}}
        \fi}

\newcommand{\tcaption}[1]{
        \refstepcounter{table}
        \setbox\@tempboxa = \hbox{\eightrm Table~\thetable. #1}
        \ifdim \wd\@tempboxa > 5in
           {\begin{center}
        \parbox{5in}{\eightrm\smalllineskip Table~\thetable. #1 }
            \end{center}}
        \else
             {\begin{center}
             {\eightrm Table~\thetable. #1}
              \end{center}}
        \fi}

\def\@citex[#1]#2{\if@filesw\immediate\write\@auxout	
	{\string\citation{#2}}\fi			
\def\@citea{}\@cite{\@for\@citeb:=#2\do			
	{\@citea\def\@citea{,}\@ifundefined		
	{b@\@citeb}{{\bf ?}\@warning
	{Citation `\@citeb' on page \thepage \space undefined}}
	{\csname b@\@citeb\endcsname}}}{#1}}

\newif\if@cghi
\def\cite{\@cghitrue\@ifnextchar [{\@tempswatrue
	\@citex}{\@tempswafalse\@citex[]}}
\def\citelow{\@cghifalse\@ifnextchar [{\@tempswatrue
	\@citex}{\@tempswafalse\@citex[]}}
\def\@cite#1#2{{$\null^{#1}$\if@tempswa\typeout
	{IJCGA warning: optional citation argument
	ignored: `#2'} \fi}}

\def\pmb#1{\setbox0=\hbox{#1}
	\kern-.025em\copy0\kern-\wd0
	\kern.05em\copy0\kern-\wd0
	\kern-.025em\raise.0433em\box0}

\def\fnt#1#2{\footnotetext{\kern-.3em
	{$^{\mbox{\scriptsize #1}}$}{#2}}}

\def\fpage#1{\begingroup
\voffset=.3in
\thispagestyle{empty}\begin{table}[b]\centerline{\footnotesize #1}
	\end{table}\endgroup}

\def\runninghead#1#2{\pagestyle{myheadings}
\markboth{{\eightit{\quad #1}}\hfill}{\hfill{\eightit{#2\quad}}}}

\font\tenbf=cmbx10
\font\tenit=cmti10
\font\tenit=cmti10
\font\bfit=cmbxti10 at 10pt
 1
 1
 1

\font\ninerm=cmr9

\font\eightrm=cmr8
\font\eightit=cmti8

\def\qed{\hbox{${\vcenter{\vbox{                          
   \hrule height 0.4pt\hbox{\vrule width 0.4pt height 6pt
   \kern5pt\vrule width 0.4pt}\hrule height 0.4pt}}}$}}

\runninghead{\it Daniel Dom\'{\i}nguez and Jorge V. Jos\' e}
{Non-equilibrium coherent vortex states ... }
\begin{document}
\normalsize\textlineskip
{\thispagestyle{empty}
\setcounter{page}{1}
\renewcommand{\thefootnote}{\fnsymbol{footnote}} 
\copyrightheading{to be published as a ``Review''
--- cond-mat}
\vspace*{0.88truein}
\fpage{1}

\centerline{\bf NON-EQUILIBRIUM COHERENT VORTEX STATES}
\vspace*{0.035truein}
\centerline{\bf AND SUBHARMONIC GIANT SHAPIRO STEPS }
\vspace*{0.035truein}
\centerline{\bf IN JOSEPHSON JUNCTION ARRAYS }
\vspace{0.37truein}
\centerline{\footnotesize Daniel Dom\'{\i}nguez$^a$ and Jorge V.
Jos\'e$^{b,c}$}
\vspace*{0.015truein}
\centerline{\footnotesize\it $^a$
Theoretical Division, Los Alamos National Laboratory,}
\baselineskip=10pt
\centerline{\footnotesize\it MS B262, Los Alamos, New Mexico 87545, USA}
\baselineskip=10pt
\centerline{\footnotesize\it $^b$Instituut voor Theoretische Fysica,
Universiteit Utrecht}
\baselineskip=10pt
\centerline{\footnotesize\it 3508 TA Utrecht, The Netherlands}
\baselineskip=10pt
\centerline{\footnotesize\it $^c$Department of Physics, Northeastern
University}
\baselineskip=10pt
\centerline{\footnotesize\it Boston, Massachusetts 02115, USA }
\vspace{0.225truein}
\pub{June 1994}
\vspace*{0.21truein}
\abstracts{\noindent
This is a review of recent work on the dynamic response of Josephson
junction arrays driven by dc and ac currents. The arrays
are modeled by the resistively shunted Josephson junction model, appropriate
for proximity effect junctions, including self-induced magnetic fields
as well as disorder.  The relevance of the self-induced fields is measured
as a function of a parameter $\kappa=\lambda_L/a$, with $\lambda_L$ the
London penetration depth of the arrays, and $a$ the lattice spacing.
The transition from Type II
($\kappa>1$) to Type I ($\kappa <1$) behavior is studied in detail. We
compare the results for models with self, self+nearest-neighbor, and
full  inductance matrices. In the $\kappa=\infty$ limit,
we find that when the initial state has at least one vortex-antivortex pair,
after a characteristic transient time these vortices
unbind and {\it radiate}  other vortices.
These radiated vortices settle
into a parity-broken, time-periodic, {\em axisymmetric coherent vortex
state} (ACVS), characterized by alternate rows of positive and negative
vortices lying along a
tilted axis. The ACVS produces subharmonic steps in the current voltage (IV)
characteristics, typical of giant Shapiro steps. For finite $\kappa$ we find
that the IV's show subharmonic giant Shapiro steps, even
at zero external magnetic field.  We find that these subharmonic steps
are produced by a whole family of coherent vortex oscillating patterns, with
their structure changing as a function of $\kappa$.
In general, we find that these patterns are due to
a break down of translational invariance  produced, for example,
by disorder or antisymmetric edge-fields. The zero field case results are
in good qualitative agreement with experiments in Nb-Au-Nb arrays.
}{}{}
\vspace*{-3pt}\textlineskip
\section{\bf INTRODUCTION}
\noindent

\noindent
There has been considerable contemporary interest in the study
of the dynamic response of two-dimensional Josephson junction arrays (JJA)
subjected to external probes $^{1-27}$. This attention has
been due in part to recent advances in photolithographic fabrication
of these arrays with tailor-made properties. The dynamics of the JJA
are described in terms of a large set of coupled, driven, non-linear
differential equations. When the current drive has dc+ac components,
novel non-equilibrium stationary coherent vortex states may appear
in the arrays.
Experimentally, giant Shapiro steps (GSS) have been observed in the IV
characteristics of  proximity-effect JJA in zero \cite{gshap1} and rational
magnetic field frustrations \cite{benz}. The frustration, $f=\Phi/\Phi_0$,
is defined as the average applied magnetic flux per plaquette,
$\Phi$, measured  in units of the flux quantum $\Phi_0=h/{2e}$.
Coherent oscillations of ground state field-induced vortices are believed
to be responsible for the existence of the {\it fractional} GSS
when $f=p/q$, with $p$ and $q$ relative primes \cite{benz}.
This interpretation was successfully verified in numerical
simulation modelling of the  arrays by a resistively shunted junction
(RSJ) model \cite{numshap1,numshap2}, as well as from analytic
studies \cite{teorshap}.
There have also been some experimental \cite{lidia} and theoretical
studies \cite{tom,holand2,chu} of the relevance of the geometry of
the arrays and the  direction of the  input current on the generation of
fractional GSS.
Moreover, the experiments in GSS have encouraged
investigations of JJA as coherent radiation sources \cite{coher}.

Later experiments on Nb-Au-Nb \cite{cinci,garland,garland2} and
Nb-Cu-Nb\cite{benz}$^{(b)}$ JJA also showed half-integer GSS
(and also some evidence for
higher order subharmonic steps) in $\underline{{\it zero}}$ magnetic field.
These steps can  not be explained within the context of the non-inductive
RSJ model, which proved successful in providing an understanding of the
$f=p/q$ Shapiro steps.\cite{numshap1,numshap2,teorshap,tom,holand2}
In this review we will show that these extra subharmonic steps can be produced
by breaking the  translational invariance in the arrays.

For example, we found  that the addition
of disorder can in fact induce extra half-integer steps in zero
magnetic field \cite{nos,nos2,nos2b} within the RSJ model, even in the limit
when no self-field effects are included.
These extra steps were found to be related to a coherent oscillation of
current-nucleated vortices. Previously,  the nucleation
of vortex-antivortex pairs by the defects, induced by an applied  dc current,
was discussed by Xia and Leath \cite{leath}.
In our studies we added an ac component to the dc current.
The addition of the ac current changes the vortex dynamics in a fundamental
way: It leads to the formation of a novel, far from equilibrium,
{\em axisymmetric coherent vortex state} (ACVS). The ACVS corresponds
to an oscillating pattern of tilted positive and negative vortex rows
which  produce the extra
half-integer GSS (see Figs 6 and 7).  The relevance of the disorder
is to act as  nucleating centers of vortex pairs, thus providing the
relevant initial conditions in the  dynamics to generate the ACVS.
However, as we shall discuss in this review, what matters in generating the
ACVS is to have a mechanism to produce vortex-antivortex pairs
(VAP) in the initial conditions since then the
dynamics are such that the ACVS is generated in most cases.
We will discuss in more detail the properties of the ACVS in Section 3.

Most of the experiments in GSS have been carried out in
proximity effect arrays, that have strong temperature-dependent critical
currents \cite{benz,cinci,garland,garland2}.
In this case self-induced magnetic field
(SIMF) effects, which also break translational invariance due to the
current induced edge-fields,
can be of relevance for the interpretation of the experimental
data as was found in Ref.~\citelow{cinci}. We have recently developed a
dynamic model to study the SIMF effects in Josephson
junction arrays\cite{nos3,nos3b,nos4}. When applied to the study of dc+ac
driven JJA, we found that
the fractional GSS observed at nonzero magnetic fields are affected by
SIMF effects. There are
two extreme regimes of interest as a function of $\kappa=\lambda_L/a$
in which the underlying microscopic coherent vortex states
responsible for the fractional GSS are qualitatively different. Here
$\lambda_L$ is the London penetration depth
and $a$ the lattice spacing.

Also, in the case of zero external field we found  subharmonic GSS, produced
by a family of oscillating coherent vortex states, with different structures
as a function of $\kappa$ \cite{nos4}. This connects with the
ACVS which turns out
to be the $\kappa\rightarrow\infty$ limit of this family of vortex states.

In this review we present  a brief recap of our main results
pertaining to the generation of coherent vortex states, mostly in
zero external magnetic field. In this way we present
 a unified view  of our understanding of
the physics of subharmonic Giant Shapiro Steps
at $f=0$, with zero and finite screening current effects.
In our initial calculations\cite{nos4} we considered the cases where
the inductance matrix had diagonal and nearest neighbor components.
This approximation has been significantly improved  by including the full
inductance matrix in the analysis of
Refs~\citelow{phillips,phillips2,majhofer2}.
Following the general idea of this improvement we have redone some of our
 previous calculations  including the full inductance
matrix and we include a critical comparison between the results of the
different approximations.

\section{\bf DYNAMICS OF NON-INDUCTIVE JJA}

\subsection{\bf The Josephson effect}

We start by briefly reviewing the essentials of the Josephson effect.
A Josephson junction is made of two small superconductors separated by a thin
film of non-superconducting material. The two basic equations that describe
the physics of the junction are
\begin{equation}
I_J=I_0\sin(\theta_2-\theta_1)\;,
\label{eq:jose1}
\end{equation}
that gives the supercurrent that flows between the two superconductors,
and the voltage drop between them  given by
\begin{equation}
V=\frac{\Phi_0}{2\pi}\frac{d}{dt}(\theta_2-\theta_1)\;.
\label{eq:jose2}
\end{equation}
 Here  $\theta_1, \theta_2$ are the phases of the Ginzburg-Landau
order parameter in superconductors $1$ and $2$,
respectively. The critical current $I_0$ is the maximum
current that can flow through the junction, and $\Phi_0=h/2e$ is the
superconducting quantum of flux. In the presence of a magnetic field
the phase difference is
replaced by its gauge invariant form, $(\theta_2-\theta_1)\rightarrow
(\theta_2-\theta_1 -\frac{2\pi}{\Phi_0}\int_1^2\vec A\cdot d\vec l)$,
where $\nabla \times \vec A=\vec B$ is the magnetic field.
In real junctions
there is always some dissipation. This dissipative effect
has been successfully modeled, in the case of proximity effect junctions,
 by the McCumber-Steward
resistively shunted junction (RSJ) model
\cite{mcumber}. In the RSJ model, for a dc current biased junction, the
total current $I$ that flows in parallel with the ideal Josephson current is
\begin{equation}
I=I_J+V/{\cal R}=I_0\sin(\Delta\theta)+
\frac{\Phi_0}{2\pi{\cal R}}\frac{d\Delta\theta}{dt}\;,
\label{eq:rsj}
\end{equation}
where we have used Eqs.~(\ref{eq:jose1}) and (\ref{eq:jose2})
and ${\cal R}$ is the RSJ shunt resistance. When
$I$ is time-independent, Eq.(\ref{eq:rsj}) can be
solved analytically.  In this case, the time averaged voltage
$\langle V\rangle$ as
a function of $I$, which defines the IV curve of the junction,
is given by $\langle V\rangle=0$ when $I<I_0 $ and
$\langle V\rangle={\cal R}\sqrt{I^2-I_0^2}$ when $I>I_0$.

When the junction is driven by a time-periodic current
$I=I_{dc}+I_{ac}\sin(2\pi\nu t)$, the $\langle V\rangle$ vs $I_{dc}$ curve
shows plateaus at the quantized voltages
\begin{equation}
\langle V_n\rangle = n\frac{h\nu}{2e},\;\;\;\;n=1,2,3,\ldots\;\;.
\label{eq:sha1}
\end{equation}
These are the Shapiro steps that have allowed very precise
measurements of the voltage unit \cite{shapiro}. The central
question addressed in this paper is what happens when we couple
a large number of Josephson
junctions to an array that is then driven by dc+ac currents.

\subsection{\bf Josephson junction arrays}

A Josephson junction array (JJA) is made of an $N\times N$  network of
superconducting islands connected by Josephson currents
\cite{korea,review2}.
For example, the square arrays of Benz {\it et al.} \cite{benz} were
made of $1000\times 1000$ Nb-Cu-Nb proximity effect junctions with a
lattice constant of $a=10\,\mu m$. A schematic
representation of a JJA driven by an external current is shown in Fig.~1.

To model the dynamical behavior of this
system we extend the RSJ model to a square JJA network
\cite{dyna}.
The current $I_{\mu }(r)$, along the $\mu $ direction between the
superconducting
islands at sites $r$ and $r + \mu $, with $\mu=\hat{e}_x,\hat{e}_y$,
is given by
\begin{equation}
I_{\mu }(r) = I_0\,\sin ({\Delta }_\mu \theta (r) - A_{\mu }(r) )\,
+\,
{{\Phi _0} \over {2\pi{\cal R}}}\,{d\over {dt}}\,(
{\Delta }_\mu \theta (r) - A_{\mu}(r))
\, +\, {\eta}_{\mu}(r,t)\;, \label{eq:rsj2}
\end{equation}
where $\Delta_\mu\theta (r)=\theta(r+\mu)-\theta(r)$. The
external magnetic field produces the frustration $f=\frac{Ha^2}{\Phi_0}$,
that measures the average number of flux quanta per unit cell, and defined by
\begin{equation}
2\pi f\,=A_x(r)+A_y(r+\hat{e}_x)-A_x(r+\hat{e}_y)-A_y(r)\equiv\Delta_\mu\times
A_\mu(r),
\label{eq:amu}
\end{equation}
 where  the link variable
$A_{\mu}(r) = \frac{2\pi}{\phi_0}\int_r^{r+\mu}\,\vec A\cdot d\vec l$.
 Here, for the moment, we are neglecting the screening currents  by assuming
that
$A_\mu(r)$ is fully determined by the external magnetic field $H$.
This assumption is correct whenever $\lambda_L \gg Na$.
For the same reason, $\frac{dA_\mu}{dt}$ can be dropped from
Eq.~(\ref{eq:rsj2})
since the magnetic field is constant in time.
In Sec.~4 we shall present the full analysis including screening current
effects. The effects of temperature are included by adding the Gaussian
random variable $\eta _\mu (r,t)$ to the equations of motion
with covariance
\begin{equation}
\langle\eta _\mu (r)\, \eta _{\mu '}(r')\rangle = {2kT\over {\cal R}}
\delta _{r,r'} \delta _{\mu ,\mu '} \delta (t-t').
\end{equation}
Eq.~(\ref{eq:rsj2}) together with
Kirchoff's current conservation law,
\begin{equation}
\Delta _\mu \cdot I_\mu (r) = I_x(r) - I_x(r-x) + I_y(r) -I_y(r-y) = I^{ext}
(r)\label{eq:kir},
\end{equation}
 valid at each node, fully define the evolution of the
phase $\theta (r,t)$ as a function of time. Here $I^{ext}(r)$ denotes the
external current injected at site $r$. The explicit expression
for  $\frac{d\theta}{dt}$ derived from these equations is \cite{dyna}
\begin{equation}
\frac{d\theta (r,t)}{dt} = -\frac{2\pi {\cal R}}{\Phi_0}\sum_{r'} G(r,r')
\{ I^{ext}(r,t) - \Delta_\mu \cdot
[I_0\sin(\Delta_\mu \theta(r',t) - A_\mu (r')) + \eta_\mu
(r',t)]\},
\label{green}
\end{equation}
with $G(r,r')$ being the two-dimensional lattice Green
function which depends on the boundary conditions chosen.
Eq.(\ref{green}) defines the  set of coupled nonlinear
dynamical equations studied in this paper. In Benz's experiment \cite{benz}
this would represent a set of
$10^6$ non-linear coupled oscillators driven by external currents.

Of course, solving this large set of coupled non-linear equations
analytically is out of the question. We use then an efficient algorithm to
study  Eq.(\ref{green}) numerically.  In our simulations
the external currents are injected uniformly along the $y$-direction with
$I^{ext}(y=0)=I(t)$
and zero elsewhere. The potentials at $y=N_Y$ are fixed to zero
by setting the phases $\theta (y=N_Y+1) = 0$ (this boundary condition
removes a singularity present in the Green function matrix). We choose
periodic boundary conditions (PBC) along the $x$-direction
in most calculations, but we have also used free end boundary
conditions (FBC) for comparison purposes.

A direct numerical evaluation of Eq.~(\ref{green})
grows as $N^4$, which is too slow and limits the sizes of
the lattices that can be simulated. Overcoming this restriction turns out to
be essential for the formation of the ACVS state. We have
used instead a very efficient method \cite{nos,nos2,nos2b} based on the use of
the Fast Fourier Transform (FFT). Specifically,  this involves
doing a FFT of Eq.~(\ref{green}) along the
$x$-direction and then solving the resulting tridiagonal matrix
equation along the $y$-direction. This algorithm grows as $N^2 \log N$,
which is appreciably faster than the direct method. Eikmans and van
Himbergen \cite{holand} were the first ones to use the FFT
along both the $x$ and $y$ directions to study JJA.
Our method shows an improvement
over theirs of about $30\%$. The time integration is carried out using
a fixed step fourth order Runge-Kutta (RK) method. Furthermore,
at finite temperatures we use an extension of the  second order RK method
for stochastic differential equations developed by Helfand and
Greenside \cite{helfand}.
Typical integration steps used were $\Delta t\nu_0 = 0.01-0.1 $,
with characteristic Josephson
frequency $\nu_0=\frac{2\pi {\cal R}I_0}{\Phi_0}$.

\subsection{\bf Periodic square arrays and giant Shapiro steps}

For ordered arrays, Eq.(\ref{green}) is invariant
against the transformations
\begin{equation}
f \to {f+n} \,\,\,\,\,\,\,\hbox{and}\,\,\,\,\,\, f\to -f.
\end{equation}
with $n$ an integer. In this case it is enough to analyze the
properties of the model in the interval $f=[0,1/2]$.
The phase vortex excitations that can appear in the JJA are
defined by
\begin{equation}
\sum_{R}\kern-1.5em\bigcirc\kern1.5em \left[\Delta_\mu\theta (r)-A_\mu(r)
\right]= 2\pi(n(R)-f),
\label{eq:vort}
\end{equation}
where the sum is over the plaquette $R$ and the gauge invariant
phase differences, $\Delta_\mu\theta (r)-A_\mu(r)$,
 are restricted to the interval $[-\pi,\pi]$. Therefore,
$n(R)$ is an integer that gives the vorticity
at plaquette $R$. In zero field and at finite temperatures vortices play a
central role in triggering the Berezinskii-Kosterlitz-Thouless
phase transition \cite{BKT}. For fractional frustrations $f=p/q$,
with $p$ and $q$ relative primes,
the ground states of the arrays consist of superlattices of field-induced
vortices with unit cell of size  $q\times q$
\cite{gstate}. These states lead to non-monotonic behavior of
the magnetoresistance and  $T_c(f)$ studied both experimentally and
theoretically\cite{korea,gstate,review2}.

We can calculate the experimentally measurable IV characteristics
from the solutions to Eq.~(\ref{green}). The calculations give the
time-averaged total voltage drop $\langle V\rangle$
per row in the array as a function
of the  applied dc current.
For periodic arrays, the $f=0$ dynamic response of the model is equivalent to
the superposition of $N$ individual Josephson junctions along the direction of
the current\cite{dyna} for
the total current flows along the y-direction. The model behavior is
that of $N_x$ rows with  $N_y$  junctions in series. Thus, when the $I_{ac}=0$,
the IV characteristics are simply given by the one-junction result
multiplied by  $N_y$. The same is true in a JJA
when $f=0$ and  $I(t) = I_{dc} + I_{ac}\sin (2\pi\nu_a t)$,
leading to {\em giant} Shapiro steps at
voltages \cite{gshap1}:
\begin{equation}
V_n=N_y\frac{nh\nu}{2e},\;\; n=1,2,3,\ldots\; .
\label{eq:gshap}
\end{equation}
The $N_y$ factor is important in the possible practical applications of JJA.
The symmetry that allows us to separate the $N_x\times N_y$ array
into  $N_x$ independent rows in series is broken when the external
magnetic field is nonzero. No exact analytic
solution to Eq.~(\ref{green}) that gives the IV characteristics is known
in this case. It was first found experimentally
\cite{benz} that there are fractional GSS at voltages
\begin{equation}
V_{n,q}=\frac{N_y}{q}\frac{nh\nu}{2e},\;\; n,q=1,2,3,\dots\; ,
\label{eq:fgss}
\end{equation}
when $f = p/q$. These fractional steps were explained as due to the collective
oscillations between different $f=p/q$ vortex ground state configurations
\cite{benz,numshap1,numshap2,teorshap}. The vortex-superlattice
oscillates in synchrony with the ac current with frequency $\nu/q$.
We have in fact seen these ground state oscillations in animations
produced with solutions to Eq.~(\ref{green}).

In the zero field case one could expect that by breaking translational
invariance, which allows the reduction of the 2-D array into effectively
1-D rows, one may get extra subharmonic GSS even in zero magnetic field.
One can  break translational invariance by simply adding
disorder to the array. This is the topic of the next section.
Later, in Sec.~5, we will see that current induced magnetic fields also break
translational invariance.

\subsection{\bf Defect nucleated vortex pairs.}

We introduce disorder in a square JJA by displacing the lattice sites radially
away from their periodic positions by a distance $\delta <1$, and
then choosing the angle from a uniform probability distribution in
$[0,2\pi]$. We assume that the critical currents
and shunt resistances are all the same and equal to
$I_0$ and  ${\cal R}$'s, respectively. This means that
the disorder is only evident when the link variables $A_\mu (r)\neq 0$.
Adding one defect to the array affects only
the four $A_\mu$'s  connecting the bonds linked to the displaced lattice site.
Varying $f$ can be seen as increasing the deformation or disorder in the
lattice, since the $A_\mu$'s are proportional to the external field.
When we take $f=n$ an integer, because of the local periodicity of the
equations  of motion with respect to $f$,
only the plaquettes around the defect will be modified by
the presence of the field. The $n$ value chosen will represent the ``strength"
of the defect. As we shall discuss later the specific nature of the
disorder will not be important in producing an ACVS.
For example, the ACVS also appears  when we cut bonds with $f=0$.

Leath and Xia \cite{leath} studied the response of a JJA with a rectangular
defect and in the presence of a constant dc current.
They found that there is a tendency to create a vortex-antivortex pair (VAP)
pinned to the defect for $I_{dc}\neq 0$.
When the external current is larger than the critical current of the array,
$I_{c}$, the Lorentz force acting on the VAP is enough
to break them apart and they can move away from the defect.
As they move along the direction perpendicular to the external current,
they produce a dissipative Faraday voltage that leads
to a nonzero resistance in the array.
A larger number of VAP's is generated
when  $I_{dc}\gg I_c$, leading to rather complex vortex motions in
the array.

In Fig.~2 we show an example of this situation for the kind of topological
defect we defined above. We show the vortex distribution
$n(R)$ for a $40\times 40$ array, which was the typical size studied,
although we will discuss results for larger lattices as well.
For $I<I_c$ we see a VAP pinned at the defect.
When $I>I_c$, vortices and antivortices move away from
the defect as found in Ref.~\citelow{leath}.

\section{\bf AXISYMMETRIC COHERENT VORTEX STATES (ACVS)}

\subsection{\bf IV characteristics}

In this section we discuss the IV characteristics of JJA with the type of
defect introduced in the previous section and driven by a current
\begin{equation}
I^{ext}=I_{dc}+I_{ac}\sin (2\pi \nu t).
\end{equation}
Let us start with a JJA with only one defect.
Later we will discuss what happens when the JJA has more defects.

In Fig.~3(a) we show the IV characteristics for a
periodic JJA. In calculating the IV characteristics as we vary
$I_{dc}$, we take  the final phase configurations of the preceding current
value  as initial conditions for the next one.
In Fig.~3(a) we see the expected integer GSS (Eq.~\ref{eq:gshap}),
which are identical to those obtained  with one Josephson junction, except
for a factor of $N_y$. In Fig.~3(b) we show the results for the
IV of a $40\times 40$ square lattice  with one defect located at the
center of the array.  We can clearly distinguish the new half-integer steps
in the IV curve at voltages
\begin{equation}
V_{n/2}=\frac{n}{2}\frac{N_yh\nu}{2e};\;\;\;\;\; n=1,2,\ldots\;\;,
\end{equation}
in addition to the expected rounding of the usual integer steps
due to the presence of disorder. It turns out
that the precise location of the defect on the lattice
has no effect on the nature of the final stationary state nor on the
appearance of the half-integer steps. The half-integer steps are found to be
hysteretic. This is seen
in Fig.~3 (c) where we show a  blow-up of the 1/2-step region,
but for an $80\times 80$ lattice.
For the 1/2-step we can define two critical currents,
$I_c^+$ when increasing the current, and $I_c^-$ when decreasing it.
As usual, the size of the hysteresis loop is history-dependent.
For $I>I_c^+$, the 1/2-step is reversible, and it can
be reached independent of the initial conditions. For $I_c^-<I<I_c^+$
the response depends directly on the initial conditions. Note the appearance
of a $2/3$-step at
\begin{equation}
V_{2/3}=\frac{2}{3}\frac{N_yh\nu}{2e},
\end{equation}
which was not clearly evident in the $40\times 40$ lattice calculation.
The $2/3$-step is also found to be hysteretic. We will discuss
this step in more detail in Sec.~3.3.

An important property of the half-integer steps is that they are not
clearly visible for lattice sizes smaller than about $18\times 18$.
We have then carried out a finite size analysis of the 1/2-step
width \cite{nos,nos2b}  and found that
there is a minimum critical size $N_c\sim 16$,  above which we start to see
indications of the existence of the 1/2-step. The 1/2-step width reaches
an asymptotic
plateau for lattices larger than about $32\times 32$.
We believe that the need to have a minimum lattice size to see the 1/2-step
is related to the specific nature of the ACVS structures that
needed  to be stable in the non-equilibrium stationary state. We discuss
this point further in the next section.

We have also calculated the IV's for JJA with free boundary
conditions. We find that the subharmonic steps are also present,
but the lattice sizes needed to see the ACVS are still larger. This is
easy to understand since
the boundary has more of a perturbing effect in smaller lattices.
As the lattice size increases the shape of the subharmonic steps approaches
the sharpness of the ones found when using PBC \cite{nos2,nos2b}. Therefore, we
conclude that the boundary conditions are not essential for the existence of
the ACVS.

Once we had identified the subharmonic steps due to the presence of
defects we studied their quantitative properties. Following the approach
used for one junction, we began the characterization  of the
half-integer Shapiro steps by studying their width-dependence
on the frequency $\nu$, the ac current $I_{ac}$, the amount of disorder
and temperature. In Fig.~4(a) we show the dependence of the
1/2-step width as a function of
$\nu $ with $I_{ac}=I_0$. Here we note that there is a frequency window
for which the 1/2-step is clearly visible and has a well defined
maximum at $\nu_{max}
\approx 0.22\, \nu_0$ \cite{nota}. The width of the frequency  window
is $\Delta \nu\equiv \nu_{+}-\nu_{-}\sim 0.043\nu_0-0.4\nu_0$. Outside this
frequency window the step width is essentially zero.

In the one junction case the step width as a function of $I_{ac}$ goes like
$\Delta I_n=2I_0J_n(\frac{2e{\cal R}I_{ac}}{h\nu})$, with $J_n$ the Bessel
function of integer order \cite{bessel}.
In Fig.~4(b) we show the oscillatory behavior of $\Delta I_{1/2}$
as a function of $I_{ac}$, for $\nu =0.1 \, \nu_0$, and with one defect.
The oscillations do not appear to fit the expected Bessel function form
but the data is not good enough to make a definitive statement. Even more
important is the fact that we appear to find a critical value for
$I_{ac}$ above which the step appears \cite{nota2}.

We need to consider temperature effects to make comparisons with the
experimental data. We have then studied the stability of the 1/2-step
as a function of temperature \cite{nos,nos2,nos2b}. We found that the 1/2-step
gets rounded as temperature increases and the hysteresis step width is reduced.
The 1/2-step tends to disappear for a relatively low temperature
$T > 0.04{\Phi_0I_0}/{2\pi k_B}$, which is small
compared to the  BKT critical temperature ($\sim O(1)$) \cite{BKT}.
This is consistent with the fact that the width of the
1/2-step is small when compared to the critical current of the array.

We also found that when we added more disorder, by increase the number of
lattice sites displaced from their periodic positions in the array,
the half-steps are still there but their hysteretic properties
get quantitatively modified\cite{nos2,nos2b}. Even
when all the lattice sites are deformed,
the 1/2-integer step is still there, but it is more rounded and with the
hysteresis loop reduced in size. We conclude that
a completely distorted lattice is not enough to destroy the
extra 1/2-steps. However, a study of the half-integer
steps as a function of the strength of the disorder \cite{nos2b}
shows that there is a critical value above which the  1/2-steps do
disappear.
This value coincides with the disorder strength above which there
is a plastic flow of vortices in dc driven JJA \cite{plastic}.

\subsection{\bf Coherent Vortex oscillating states}

In the previous section we described the  conditions
under which extra subharmonic  steps appear in the IV characteristics
in the $\kappa =\infty$ limit.
Here we want to discuss the physical origin of these steps based
on a microscopic analysis of the vortex dynamics.

A useful quantity to look at is the spectral function
\begin{equation}
S(\nu)=\lim_{\tau
\rightarrow \infty}\vert {1\over
{\tau}}\int_0^{\tau} V(t)e^{i{2\pi\nu }t}dt\vert ^2
\end{equation}
as a function of frequency $\nu$.  We calculated $S(\nu)$ and found,
as would have been expected, that it shows resonances
at frequencies  $n\nu /2$ for the 1/2-steps \cite{nos2,nos2b}.
The resonances at the 1/2-step indicate that there is an underlying
coherent vortex oscillatory state with twice the period of the ac current.
These resonances only exist for currents right on the steps, fractional and
integer, while for other values of $I_{dc}$ there is a broad band spectrum
in $S(\nu)$.

As mentioned in Section 2.3 the presence of {\sl fractional} giant
Shapiro steps, found when $f=p/q$, is explained in terms of coherent
vortex oscillations  between different lattice ground states.
At zero field and at $T=0$ there are no vortices in a periodic array.
However, as we showed in Sec.~2.4 for the dc
case, the presence of  defects can produce current-induced vortices.
Therefore, to further study the vortex dynamics we calculated
the total absolute number of vortices in the array
\begin{equation}
N_a(t) = \sum_R \vert n(R,t)\vert,
\end{equation}
as well as the total vorticity
\begin{equation}
N_T(t)=\sum_R n(R,t)=N_++N_-,
\end{equation}
where $n(R,t)$ was defined in Eq.~(\ref{eq:vort}). We started by
choosing $I_{dc}$  within the reversible part of the 1/2-step.
Typical results for $N_a(t)$ and $N_T(t)$ in the single-defect case
are shown in Fig.~5. From this figure we can identify two characteristic
times $t_d$ and $t_{ACVS}$, the {\it dissociation time}
and the {\it ACVS time}, respectively.
For $t<t_d$ there is a vortex pair (as in the dc case, Fig.~2) but with
its $\pm$ polarity oscillating  in phase with the current.
The size of the  VAP increases with time due to the oscillating Lorentz force
felt by the vortices.
At $t_d$ the  vortex-dipole breaks up and the vortices move away from the
defect.  After that time the number of vortices increase steeply.
There is a novel mechanism that generates vortex pairs from the moving ones.
This kind of vortex ``radiation'' phenomenon appears to be catalyzed by
the presence of the oscillating field due to the presence of the dc+ac current.
The vortex motions in this intermediate phase do not follow a
regular pattern. In the one-defect case the transient state
has a left-right antisymmetry about an axis centered about the defect and
along the current direction. This antisymmetry is reflected
in $N_T$, which is exactly zero in this time interval.
After the $t_{ACVS}$ time the
antisymmetry is broken, and an {\it axisymmetric coherent vortex
state} (ACVS) starts to form. This is seen in the $N_a(t)$ curve as
periodic variation of the number of vortices, and in $N_T(t)$
as fluctuations about zero.


In Fig.~6(a) we show
the distribution of vortices and antivortices in the ACVS state.
The vortex configurations are formed by rows
of vortices of alternating sign lying along a tilted axis with an angle
of about $27^\circ$ measured clockwise with respect
to the $x$-axis. The rows are separated along the $x$-axis by a distance of
about $N_x/2$. In total there are $N_y/2$ vortices and $N_y/2$ antivortices.
This state oscillates in sign in phase with the ac current.
After a time $1/{\nu}$  the vortex rows interchange their signs, and after
$2/\nu$ they are back to the previous vortex state. Therefore, the vortex
configurations change periodically in time with a period twice that of the
ac current. During most of the time the vortex patterns correspond to one or
the other possible ACVS configuration.
The change of sign of the vortex rows happens in a very short time,
corresponding to the peaks in the $N_a(t)$ curve of
Fig.~5(a). At that instant in time the vortices and
antivortices appear to cross from one row to the other. The collisions between
them produce additional vortices and antivortices, which annihilate each
other rapidly  before a new ACVS of opposite sign is formed.

The ACVS remains stable for the largest times considered ($10^4/{\nu}$).
To study the stability of the ACVS, we turned off the ac or the dc
currents at times $t_{off}$ as shown in Fig.~5. In both cases,
after turning off one of the currents with the other one still on,
the ACVS collapses into the initial VAP pinned  state.
The total vortex number decreases via direct vortex-antivortex annihilations.
However, the process of annihilation is much slower
when the ac current is still on while the dc current is turned off.
This means that the ac current strongly affects the collisions between
vortices by actually delaying the direct annihilation of vortices by
antivortices. Within a 1/2-step the ACVS is also stable against small
changes in the frequency and magnitude of the ac current and the frequency.

A remarkable aspect of the ACVS state is that its structure,
characterized by its angle, the distance between vortex rows and
the number of vortices, appears to be a very robust
non-equilibrium fixed point attractor of the dynamics. For example,
we also have considered the cases when two, three, or all
the lattice sites are disordered. We find that in these cases,
although the quantitative values of the transients and the number of
disorder-induced vortices is different, the stationary oscillatory state
is essentially an ACVS. Moreover, there are parameter windows
for the ac current, frequency,  defect ``strength'' $f$ and temperature,
where the ACVS shows the same vortex pattern.
The periodically oscillating vortex pattern is also the same  for higher order
$n/2$-steps with period $(n+1)/{\nu}$. However, we needed larger lattices,
at least $64\times 64$, to ``fit in" an  ACVS  when the boundary conditions
were changed to FBC, although the ACVS patterns  appeared  slightly
``distorted'' from the ones  obtained using PBC. We conclude that
boundaries  can affect the coherence of the state, but for large enough
lattices the ACVS forms and is stable.

We note that the ACVS can form in the two possible broken-symmetry states
with angular orientations either $27^\circ$ or its complimentary
$180^\circ - 27^\circ$. Furthermore, we found that for large
rectangular lattices (when $N_y >> N_x$)  the two broken-symmetry states
can coexist with each other \cite{nos2,nos2b}.

At this point we wanted to know what essential ingredients are needed
to nucleate an ACVS. A naive approach would suggest that
the steady state is reached when the rate of vortex generation by the defect
is equal to the rate of vortex annihilation.
However, we found that most vortices were generated away from
the defect during the transient time ($t_d < t < t_{ACVS}$).
If disorder was essential in the formation of the
ACVS, then when we switch off the disorder the state should collapse.
However, the ACVS remained stable when we removed the defects by setting
$f=0$. We also tried switching off the defects
for times  $t_d<t<t_{ACVS}$ and the ACVS is still stable.
On the other hand, for times $t<t_d$ the state collapses after we set $f=0$.
This suggested that once we have a large enough VAP, so that the vortices
can not annihilate each other, the VAPs are then capable of generating a
whole ACVS by themselves. This proves that what is  essential in the
formation of an ACVS  is the ``radiation''
mechanism  that produces and annihilates VAPs during their
collisions.  We also checked that if we start with a VAP
in an ordered lattice, the ACVS is generated
for the same $I_{dc}$, $I_{ac}$ and $\nu$ as when a defect was present.
It appears then that:

\newpage

{\bf {\em Any physical mechanism (not only from disorder) that
is capable of producing  a vortex-antivortex pair sufficiently far
apart, can nucleate an  ACVS in a two-dimensional JJA.}}

\subsection{\bf Higher order ACVS}

We already mentioned that when increasing the size of the lattice there are
new subharmonic steps in the IV characteristics.
This is  shown in Fig.~3 (c), where an additional voltage step appears at
\begin{equation}
V_{2/3}=\frac{2}{3}\frac{N_yh\nu}{2e},
\end{equation}
for a lattice of size of $80\times 80$. We also found that this step
appears clearly after a minimum lattice size, which is about $N_c\sim 70$.
We found that  the time evolution of the vortex patterns responsible for the
 2/3-step  also has the general structure of an ACVS, but
 now oscillating with period $3/{\nu}$. This state becomes
stable only after a very complex transient with a longer $t_{ACVS}$ than
the one found in the 1/2-step case. In Fig.~6(b) we show the stationary vortex
distributions  for a current within the 2/3-step.
It consists of rows of vortices (and antivortices) lying along a
tilted axis with the same angle of about $27^\circ$ with respect to
the $x$-axis and separated by a distance of $N_x/3$ along the x-direction.
The total number of vortices is now $\frac{3}{2}N_y$. To have period
3 there is one row of vortices and two of antivortices. They
interchange positions in synchrony with the ac-current. The
total vorticity is still zero on the average, for the rows with
vortices have double the density of those with antivortices.
This implies that there may still be more possibilities for
complicated ACVS-like oscillating patterns with the
 broken symmetry states in larger lattices. We note, however, that the
 step-width of the 2/3-step is smaller than that for the 1/2-steps. One can
 speculate that the higher subharmonic states possibly present
 in larger lattices will have step-widths that will tend to zero as
 the size of the lattice tends to infinity. The IV characteristics could then
 have  a Devil's staircase type of structure.


\subsection{\bf Experimental observations of subharmonic GSS at $f=0$}

In the previous sections we have seen that fractional giant Shapiro steps,
produced by oscillating collective vortex states, can be generated either by
having a fractional frustration $f=p/q$ in periodic arrays or by
specific initial conditions that contain a vortex-antivortex pair (VAP)
in the $f=0$ case.
There we showed that the physical origin of the two types of collective vortex
states is physically very different. These two scenarios for producing
GFSS do not appear to be able to explain the experiments that show Shapiro
steps at fractions $n/2$ and $n/3$ in {\it zero magnetic
field} \cite{cinci,garland,garland2,benz}$^{(b)}$.
Since these subharmonic  steps were not observed in single
Nb-Au-Nb junctions\cite{single,cinci,garland}, they must be
due to  the non-linear coupling between junctions in the
two dimensional arrays.
Therefore, they cannot be understood  within the RSJ model given in
Eq.(\ref{green}), because in the $f=0$ case the dynamics of a periodic array
is reduced to that of an effective single  Josephson junction.
However, as we discussed in the
previous section, the ACVS produces subharmonic steps in zero field within
the RSJ model, provided there is at least one VAP in the initial conditions.
Therefore, an
ACVS could be a good candidate to explain the experimental results observed in
Ref.~\citelow{cinci,garland,garland2,benz}$^{(b)}$.
However, these experiments were done
on (nominally) ordered arrays. More importantly, as it was
recognized in the experimental papers,
the critical currents in the SNS samples were large and therefore self-induced
magnetic fields, not included in the RSJ model, could play
a role in explaining these extra steps \cite{cinci,garland}.

The experiments in GSS have been carried out, for the most part,
in proximity effect arrays that have strong temperature-dependent
critical currents \cite{benz,cinci,garland,garland2}.
A carefully controlled narrow temperature range was
studied experimentally in order to minimize  the effect of the
self-induced magnetic fields (SIMF) \cite{benz} .
For example, in Benz's experiments the value of the
London penetration depth quoted in reference 4(b) is
$\lambda_L (T_c=3.5K)=280$  (this estimate is obtained from using Pearl's
formula, which is strictly valid in the continuum limit).
Note that $\lambda _L/a <N=1000$, which is in principle out of the regime
of validity of the model given in Eq.(\ref{green})
(e.g. $\lambda_L \gg Na$).  $\lambda_L $ decreased further when the
temperature was lowered below $T_c$.
Moreover, when going down to $T=2.5K$, a zero-field 1/2-step was observed by
Benz \cite{benz}$^{(b)}$,
with a value for $\lambda_L =2.5a$, which is certainly far from the region of
applicability of the RSJ model. Also, in  the experiments
by H. C. Lee at al. \cite{cinci} a value of $\lambda_L = 2.2a$ was quoted.

\subsection{\bf Antisymmetric edge magnetic fields}

The discussion given above suggests that the SIMF can be essential for
the understanding of the experimental data. In fact, it was found in the
experiments of Ref.\cite{cinci} that the currents flowing in the array
(at $f=0$) produce an induced antisymmetric
magnetic field distribution in the arrays. The magnetic field
is stronger at the edges of  the array along the direction
parallel to the external current, with magnitude $B_{edge} \approx
+ M I^{ext}/a^2$ at one edge, and $B_{edge}\approx -M I^{ext}/a^2$ at the
other. In the center of the array the magnitude
of $B$ tends to be small. Here $M$ is the mutual
inductance between plaquettes in the array.
Note that these current induced magnetic fields, due to their
antisymmetric nature, {\it also break translational invariance in the JJA}.

A phenomenological way of trying to mimic the
{\it edge-fields}, is to add them as boundary conditions
to the non-inductive RSJ model.
 We have done numerical studies adding the edge-fields
with flux  $\Phi = \gamma I^{ext}\Phi_0$ at one edge of the array
and $\Phi = -\gamma I^{ext}\Phi_0$ at the other\cite{nos2b,nos4}.
Again we found extra 1/2-steps in the IV
characteristics for all $\gamma$ parameter values  studied
($0.05<\gamma < 2$), just as we found before in arrays with defects.
When looking at the animation of the vortex patterns responsible for
these steps we found the same type of ACVS as discussed in the previous
section. An specific example is shown in shown in Fig.~7.
This pattern oscillates with  period $2/{\nu}$ when the current is in the
1/2-step. Looking at the transient that produces this ACVS,
we found that the edge-fields produce VAPs, and then the
dc+ac currents proceed to help nucleate the ACVS state, although the transient
is longer and more complicated than in the case with defects.

This result is in agreement with the conclusion
we reached at the end of Sec.~3.2 in that what is needed
to produce an ACVS is simply a mechanism to initially nucleate VAPs
in the array. These are the initial conditions that appear to lead
to the ACVS. Note that in the calculations described here we
used FBC.

\section{\bf DYNAMICS OF INDUCTIVE JJA}

We just discussed how an ACVS can be produced by artificially imposing
the edge magnetic fields at the boundaries of an array described
by the RSJ equations of motion. However, the real antisymmetric magnetic
fields measured experimentally must be produced dynamically from Faraday's
law. In order to properly take  into account screening current
effects we must solve self-consistently the flux and phase dynamical
equations. This is the goal of this section.

Here and in Sec. 5, we will describe our results from including the SIMF
at different levels of approximation, together with
a qualitative comparison to the experimental results. As mentioned in the
introduction, our initial approximate calculations\cite{nos4}
to the inductance matrix were
recently superseded by including the full inductance matrix in the
calculations\cite{phillips2}. Here we present our way of including the
full inductance matrix into the calculation, which differs in the details
from the approach followed by Phillips et al., but that it leads to similar
physical results. We also present a comparison of our results obtained within
the different inductance matrix approximations.

\subsection{\bf Self-consistent JJA dynamical equations}

We start by noticing that the current along the bonds of the lattice,
$I_\mu(r,t)$, is  still given by
Eq.(5). The difference comes only from the fact that in this case
$$
\frac{dA_\mu(r,t)}{dt}\not= 0,
$$
where the vector potential $A_\mu(r,t)$ is now related to the {\bf total}
magnetic flux $\Phi (R,t)$ at plaquette $R$:
\begin{equation}
\frac{2\pi\Phi(R,t)}{\Phi_0}=
A_x(r,t)+A_y(r+\hat{e}_x,t)-A_x(r+\hat{e}_y,t)-A_y(r,t)=
\Delta_\mu \times A_\mu(r,t)\, .
\label{PA}
\end{equation}
In this case we see that the total $\Phi(R,t)$ now depends on the flux
generated by the external field $\Phi_x=Ha^2=f\Phi_0$, plus the
magnetic flux induced
by all the currents flowing in the array. Specifically we can write
\begin{equation}
\Phi(R,t)=\Phi_x(R)+\sum_{r',\mu'}\Gamma(R,r',\mu')I_{\mu'}(r',t)\;,
\label{PLI}
\end{equation}
where $\Gamma(R,r',\mu')$ is
a matrix that explicitly depends on the geometries of the array and the
junctions. It is convenient to write Eq.~(\ref{PLI}) only in
terms of dual or plaquette variables. From the current conservation
condition given in  Eq.~(\ref{eq:kir}), we have that
\begin{eqnarray}
I_x(r,t)&=&J(R,t)-J(R-\hat{e}_y,t)\nonumber\\
I_y(r,t)&=&J(R-\hat{e}_x,t)-J(R,t)+I_{ext}(t)\nonumber\;,
\end{eqnarray}
with $J(R,t)$  the plaquette's current, defined to be positive for
currents flowing in the anticlockwise direction. We rewrite these
equations in a short-hand notation,
\begin{equation}
I_\mu(r)=\Delta_\mu\times J(R) + \delta_{\mu,y}I_{ext}\;,
\label{IrJ}
\end{equation}
indicating that the external current is only applied along the $y$-direction.
In terms of the plaquette variables Eq.~(\ref{PLI}) can be written as,
\begin{equation}
        \Phi(R,t)=\Phi_x(R)+\sum_{R'} L(R,R')J(R',t)+E(R)I_{ext}(t)
\label{PLJ}
\end{equation}
where
\begin{equation}
L(R,R')=\Delta_{\mu'}\times\Gamma(R,r',\mu')
\label{L}
\end{equation}
is the inductance matrix of the array and we defined
\begin{equation}
E(R)I_{ext}(t)=\sum_{r'}\Gamma(R,r',\hat{e}_y)I_{ext}(t)\;.
\label{edgem}
\end{equation}
This term gives the magnetic flux induced by the applied external
currents which is antisymmetric along the $x$-direction.
These magnetic fields have maximum amplitude at the edges of the
array and decrease towards its center, and thus they are called ``edge
magnetic fields''.  For example, for a current sheet they are given by,
$$E(R_x,R_y)\approx\frac{\mu_0a}{2\pi}
\ln(\frac{N_xa-R_x}{R_x}),
$$
with $R_x\in [a,(N_x-1)a]$. This result is a good approximation
for the actual value of $E(R)$ in JJA \cite{nos3b}.
Note that the magnitude of $E$ depends
directly on the size of the array, and it decreases at the center
of the lattice as lattice  size grows.
In the limit $|\frac{N_xa}{2}-R_x|\ll N_xa$,
$E(R_x)\approx\frac{2\mu_0}{\pi N_x}(\frac{N_xa}{2}-R_x)$ at the
center of the array. We note that there are other ways of separating
and interpreting the
different contributions in Eq.(\ref{PLJ}), as in Ref.~\citelow{phillips2}.

We note that to write down the set of dynamical equations for $A_\mu(r,t)$
and $\theta (r,t)$ we need to fix a gauge.
In the Coulomb gauge, $\Delta_\mu\cdot A_\mu(r,t)=0$,
and from the combination of Eqs. (\ref{eq:rsj2}), (\ref{PA}), (\ref{IrJ}) and
(\ref{PLJ}) we obtain again Eq.~(\ref{green}) for $\theta(r,t)$, but
now complemented by the following dynamical equation for the flux
\begin{eqnarray}
\frac{d\Phi(R,t)}{dt}&=&{\cal R}I_0\Delta_\mu\times\sin(\Delta_\mu\theta(r,t)
-A_\mu(r,t))\nonumber\\
&+&{\cal R}\Delta_\mu^2\left [\sum_{R'}L^{-1}(R,R')\left
(\Phi(R',t)-\Phi_x-E(R')I_{ext}\right )\right ]+\xi(R,t)\;,
\label{flux}
\end{eqnarray}
with the noise function,
$$
\xi(R,t)={\cal R}\Delta_\mu\times\eta_\mu(r,t).
$$
The complete dynamics of the array is now governed by Eq.~(\ref{green})
and Eq.~(\ref{flux}). These equations have to be solved
for $\theta(r,t)$ and $\Phi(R,t)$ self-consistently, where
each one of these variables
has its own characteristic frequency. We can define typical values for
these frequencies from linearizing the  Josephson term in
Eq.(\ref{flux})\cite{nos3} giving,
$ \nu _{\theta}=2\pi {\cal R}I_c/{\Phi _0}=\nu_0,$
and $\nu _{\Phi}={{\cal R}/{\mu_0 a}}.$
To carry out an efficient numerical integration of the
equations in time it is essential to take into account the
relative magnitudes of the two relaxation times.
In the extreme Type II regime $\nu_\Phi \gg \nu_\theta$ and the fast decaying
variables are the fluxes while the
$\theta $'s are slow, with the opposite happening when
$\nu_\Phi \ll \nu_\theta$.
This situation is typical of `stiff' problems in ordinary differential
equations, which are notoriously
difficult to treat analytically and even numerically, for they lead to
singular perturbations \cite{recipes}. On the other hand, since the
equations of motion are gauge invariant we can use this symmetry
to find the most appropriate gauge to solve the problem. It turns out,
however, that a fixed gauge does not allow us to efficiently solve
the equations for all values of $\nu_\Phi/\nu_\theta$.
In the $\nu_\Phi/\nu_\theta\gg 1$ limit, the extreme Type II regime,
a convenient gauge to choose is the
 Coulomb gauge. We have implemented an algorithm that works in this case.
Our discussion here will concentrate on the intermediate regime
$0.05\leq \nu_\Phi/\nu_\theta \leq 20$, where the stiffness problem is
less severe. Furthermore, this is the regime considered in the experiments
of Ref. \cite{cinci,garland,garland2}, and it corresponds to the temperature
range, $[1.5\,K,2.6\,K]$ in Benz's experiments \cite{benz}$^{(b)}$.
In this $\nu_\Phi/\nu_\theta$ range
it is more convenient to use the temporal gauge
to efficiently solve the equations. This gauge has been used extensively in
the past \cite{abrikosov} and more recently\cite{majhofer} within the
context of a JJA model of  ceramic high temperature superconductors.
This gauge entails replacing
$$(\Delta_{\mu} \theta  (r,t)- A_{\mu} (r,t))\rightarrow \Psi_{\mu}(r,t).$$

In terms of $\Psi _{\mu}$ the Langevin dynamical equations of motion read,
\begin{eqnarray}
\frac{\Phi_0}{2\pi{\cal R}}\frac{d\Psi_\mu(r)}{dt}&=&
-\frac{\Phi_0}{2\pi}
\Delta_\mu\times\sum_{R'}L^{-1}(R,R')(\Delta_\mu\times\Psi_\mu(r')+
2\pi f +\frac{2\pi}{\Phi_0}E(R')I_{ext})\nonumber\\
& &-I_0\sin\Psi_\mu(r)+\delta_{\mu,y}I_{ext}-\eta_\mu(r,t)\;.
\label{temp}
\end{eqnarray}
This system of  equations describes the same physical dynamical
evolution as Eqs.~(\ref{green}) and (\ref{flux}),
(note that in both gauges there are $2N^2$ dynamical variables).
Of course, we have to take free
boundary conditions in order to allow the internal and external magnetic
fields to relax to their correct stationary values.
This implies that relatively large lattices have to
be simulated to get the physically correct asymptotic behavior.

\subsection{\bf Inductance matrix models}

A general analytic expression for the inductance matrix $L(R,R')$, valid for
arbitrary $(R,R')$ and array geometry, is not known.  However, we can
learn a lot about the main qualitative properties of the array
from the asymptotics of $L(R,R')$ and its general symmetries.
We start by writing the standard definition of the
$\Gamma(R,r',\mu')$ matrix defined in Eq~(\ref{PLI})
\begin{equation}
\Gamma(R,r',\mu')=\frac{\mu_0}{4\pi}\frac{1}{S_RS_{r'\mu'}}
\oint_R\int_{r'\mu'}
\frac{d{\vec l}_R\cdot d{\vec l}_{r'\mu'}\,dS_RdS_{r'\mu'}}
{\vert {\vec\rho}_R-{\vec\rho}_{r'\mu'}\vert}\; .
\label{indugamma}
\end{equation}
Here $S_{R}$, and $S_{r'\mu'}$ are the cross sectional areas  of the
junctions in the plaquette $R$ and the branches $r',\mu'$, respectively (
the integrals are along the links between superconducting
islands). This expression for $\Gamma$ makes explicit its dependence
on the geometrical characteristics of the array and the junctions.
Using Eq~(\ref{L}), the representation of $L(R,R')$ based on
Eq~(\ref{indugamma}), gives
\begin{equation}
L(R,R')=\frac{\mu_0}{4\pi}\frac{1}{S_RS_{R'}}
\oint_R\oint_{R'}
\frac{d{\vec l}_R\cdot d{\vec l}_{R'}\,dS_RdS_{R'}}
{\vert {\vec\rho}_R-{\vec\rho}_{R'}\vert}\; .
\label{indutotal}
\end{equation}
This expression depends on the particular shape of the junctions and geometry
of the JJA. Here we note that  the general qualitative properties of the
response of the array to external probes will not depend on the detailed
form of the full inductance matrix. The important properties of $L(R,R')$ are
that  $L(R,R')$ is positive, its non-diagonal elements are
negative and decrease rapidly at long distances, plus  the condition
$\sum_{R'=-\infty}^{\infty}L(R,R')=0$ must be  fulfilled (because
of the continuity of the flux lines).

Here we will assume that the JJA has square lattice geometry
and that each bond is made of  cylindrical wires: Then one can obtain
$L(R,R')$, in principle, from a direct integration of Eq~ (\ref{indutotal}).
This is not exactly the geometry of the arrays considered experimentally,
and we will discuss how this approximation may affect the final results.

First we note that for large distances we can approximate the lattice
problem by its continuum limit leading to
\begin{equation}
L(R,R')_{\vert R-R'\vert \gg a}
\approx -\frac{\mu_0}{4\pi}\frac{a^4}{\vert R-R'\vert^3}\; ,
\label{longdis}
\end{equation}
which corresponds to the field of a three-dimensional magnetic dipole
produced by a current loop.  Note that the long range behavior is
always given by Eq.~(\ref{longdis}) (since it is independent of the
particular shape of the junctions).

The short distance properties of the $L(R,R')$ matrix depend
more explicitly on the specific geometry of the junctions. Here we take the
results of an explicit asymptotic evaluation of $L(R,R')$
obtained by direct integration of  Eq.~(\ref{indutotal}) for a square
network of cylindrical wires\cite{pepe,majhofer2}.  For example, the local
behavior of  $L(R,R')$ is found to be given by
\begin{equation}
\begin{array}{rcccl}
L(R,R)&=&L_0&=&\frac{\mu_0a}{2\pi}\left[8\ln\frac{2a}{r(1+2\sqrt{2})}
+8\sqrt{2}-14\right]\\
 & & & &\\
L(R,R\pm\mu)&=&-M&=&-\frac{1}{4}L_0+\frac{\mu_0a}{2\pi}0.141875 \\
 & & & &\\
L(R,R\pm e_x\pm e_y)&=&-M_{11}&=&-\frac{\mu_0a}{2\pi}0.4\ldots\;,
\end{array}
\label{lrl}
\end{equation}
where $a$ is the  lattice constant and $r$ is the radius of the wires.
For the JJA we take $2r$ as the typical width  of the junctions.
{}From now on we normalize the inductance matrix by $\mu_0a$, or
$\Lambda(R,R')=L(R,R')/\mu_0a$, and
$\Lambda_0=L_0/\mu_0a$, and ${\cal M}=M/\mu_0a$.
We use $a/r=10$, which is a typical value for arrays made with SNS
junctions\cite{cinci}, for which, $\Lambda_0=1.13...$,
${\cal M}=0.14...$, ${\cal M}_{11}=0.064...$ .

We have considered three different models for the inductance matrix:

{\underline {\bf {Model A}}} is the  simplest approximation that incorporates
screening effects including only the diagonal  or self-inductance
contribution to $L(R,R')$, i.e.
\begin{equation}
L(R,R')=\Lambda_0\mu_0 a\delta_{R,R'}.
\label{self}
\end{equation}
This approximation leads to null edge magnetic fields,  $E(R)=0$ for all $R$.
Model A  has been used in the past by Nakajima and Sawada \cite{nakajima}
to study vortex motion, and by Majhofer {\it et al} \cite{majhofer} to model
irreversible properties of ceramic superconductors.
This model is good when trying to describe the properties of bulk samples,
and three-dimensional arrays\cite{3djj}.

{\underline {\bf {Model B}}} improves model A in that it includes the
nearest neighbor mutual inductance contributions:
\begin{equation}
L(R,R')=\Lambda_0\mu_0 a\delta_{R,R'}-{\cal M}\mu_0a\delta_{R,R'\pm\mu}.
\label{mut2}
\end{equation}
In this case the edge-fields as defined above are explicitly
included. The corresponding magnetic fields at the boundaries are given by
$E({\mbox{left boundary}})={\cal M}\mu_0a$, $E({\mbox{right
boundary}})=-{\cal M}\mu_0 a$, with $E(R)=0$ otherwise.
This model was introduced by us to study the effect of screening currents
on the nucleation of
giant Shapiro Steps \cite{nos4} and magnetic properties of JJA \cite{nos3}.

{\underline {\bf {Model C}}} includes the {\it full-range} inductance matrix
in the calculations. We model $L(R,R')$ in this case  by taking into
account  the local geometry of the junctions in the diagonal and mutual
inductance contributions, as specified in Eq(\ref{lrl}), using the
filamentary wire approximation for the remaining
terms.
We model $E(R)$ in this case by an evaluation of Eq.~(\ref{edgem}) in
the filamentary approximation.
The inclusion of the full-range inductance matrix  in the study of JJA
was first considered by Phillips {\it et al} \cite{phillips},
who  developed an efficient algorithm to study static vortex properties in
JJA. Also Reinel {\it et al.}\cite{majhofer2} recently implemented a
full inductance
matrix approach to calculate static properties, somewhat closer in spirit
to the method we implemented to study the dynamics response of the JJA.
More recently Phillips {\it et al.}\cite{phillips2} have extended their
static approach to study the dynamic response of JJA.
We will compare our results for models A through C to theirs  where
appropriate. We have also studied the dynamical magnetic properties of Model
C\cite{nos3b}. Here we concentrate on the study of
the coherent vortex oscillating states of model C, which lead to
screening-induced fractional giant Shapiro steps.

Note that the essential difference between models A,B and C is that the
former assumes that the magnetic field lines are constrained to lie on the
plane while the latter takes fully into account the  three dimensionality
of the physical problem.
Of course, the complexity of the algorithms needed to solve
the dynamical equations grows with the range of the inductance matrix.
We do not discuss the specific details of the implementation of these
algorithms, which can be found in  Ref.~\citelow{nos3b}, but will instead
concentrate on discussing the corresponding results for models A through C.

\subsection{\bf Linearized equation results}

Before we discuss the numerical results obtained for models A through
C we note that there is useful physical information that can be
extracted from the
linearized approximation to the JJA equations. Including screening current
effects leads to a finite London penetration depth $\lambda_L$.
The linear approximation of the Josephson term in the stationary limit
of Eq.(~\ref{flux}) leads to the London equation for $\Phi(R)$ in a
square lattice
\cite{nos3,nos3b}.  In this case one can define an effective penetration depth
that depends on the specific model for $L(R,R')$. In the local approximation
of models A and B the penetration depth is
\begin{equation}
\lambda_J=\sqrt{\frac{\Phi_0a} {2\pi I_0\mu_0\bar\Lambda}},
\label{linearA}
\end{equation}
with $\bar\Lambda$ an effective inductance,
$\bar\Lambda=\Lambda_0$ in model A and $\bar\Lambda=\Lambda_0-4{\cal M}$
in model B. Since the lattice constant $a$ plays the  role
of a coherence length in the JJA, we can define a $\kappa_J$-parameter
\begin{equation}
\kappa_J=\frac{\lambda_J}{a}=\sqrt{\frac{\Phi_0}{2\pi
I_0\mu_0a\bar\Lambda}}=\sqrt{\frac{\nu_\Phi}{\nu_\theta\bar\Lambda}}.
\label{kappaj}
\end{equation}
For finite range inductance matrices, one can
calculate a demagnetization factor $d$ in the array.
Here $d$ is defined by the difference between the external and internal
magnetic fields $\Delta H=-d {M}$ with $d=-{1\over {L(R,R) }}\sum _{R' \neq R}
L(R,R')$, and the total magnetization ${M}$ is proportional to
$\sum _{R}J(R)$. In an external field
the array behaves as if having an effective self-inductance $\bar \Lambda =
(1-d)\Lambda_0$. In the diagonal case $d=0$, while in model B,
$d= 4{\cal M}/{\Lambda_0}$. In the thermodynamic limit $d\to 1$.
One can  show\cite{nos3b}  that, within the linearized local approximations,
for models A and B the effective vortex-vortex interaction energy
(neglecting the lattice pinning potential) goes like
\begin{equation}
{\cal U}(\rho)=\pi E_J[\ln{\frac{\lambda_J}{\rho}}+\ln
2 -\gamma]\;\;\;\; \mbox{for} \;\;\rho\ll\lambda_J,
\label{inteab}
\end{equation}
with $\gamma$ Euler's constant and
\begin{equation}
{\cal U}(\rho)=\pi E_J\frac{e^{-\rho/\lambda_J}}{\sqrt{2\pi\rho/\lambda_J}}
\;\;\;\; \mbox{for} \;\; \rho\gg\lambda_J
\label{inteab2}
\end{equation}
with $\rho=\vert R-R'\vert$ the distance
between two vortices, and the Josephson energy $E_J=\frac{\Phi_0I_0}{2\pi}$.

As was found originally by Pearl \cite{pearl} for thin films,
when considering model C, the long range screening
currents affect the effective penetration depth and the long range
vortex-vortex interactions. From linearizing Eq.(~\ref{flux}), in the static
limit, one gets\cite{nos3b,phillips,lat}
\begin{equation}
\lambda_p=\frac{\Phi_0}{2\pi I_0\mu_0}
\label{lambdap}
\end{equation}
with the corresponding $\kappa$-parameter,
\begin{equation}
	\kappa_p=\frac{\lambda_p}{a}=
\frac{\Phi_0}{2\pi I_0\mu_0a}=\frac{\nu_\Phi}{\nu_\theta}\;.
\label{kappap}
\end{equation}
In this case the effective vortex-vortex interaction energy is\cite{nos3b},
\begin{equation}
{\cal U}(\rho)=\pi E_J[\ln{\frac{\lambda_p}{\rho}}+\ln 2 -\gamma]
\;\;\;\; \mbox{for} \;\;\rho\ll\lambda_p,
\label{fullu}
\end{equation}
 while
\begin{equation}
{\cal U}(\rho)=\pi E_J \frac{\lambda_p}{\rho}
\;\;\;\; \mbox{for} \;\;\rho\gg\lambda_p.
\label{long}
\end{equation}
Here we see that ${\cal U}(\rho)$ has a slower decrease with
$\rho$ than the one given in Eq.(~\ref{inteab2}). This $\rho$ dependence
was found in the static numerical calculations of Phillips
 {\it et al.} \cite{phillips}, and it is equivalent to the one found in thin
films \cite{pearl}.

We see then that the main difference between models A and B, in the way we
have defined them,  is that
model A does not include edge magnetic fields and model B does,
but both have the same long range vortex-vortex interactions.
In contrast, although models B and C both include edge magnetic fields,
their corresponding  long range vortex-vortex interactions are different.

As we shall discuss further in this review, by considering models A through C
we can study the effects of having local screening, long-range
screening, and antisymmetric edge magnetic fields independently.
Of course in real experimental samples, in principle,
it is the full inductance matrix
plus the antisymmetric edge magnetic fields
that we should take into account as
emphasized by Phillips {\it et al.}~\cite{phillips2}.

In this review we concentrate  on the dynamic properties
of inductive JJA. It is of interest, however, to mention here the results we
have also obtained for the static  magnetic and transport properties
of inductive JJA as a  function of $\kappa$ (either $\kappa_J$ or
$\kappa_p$) for models A through C \cite{nos3,nos3b}.
We can separate the results
between extreme Type I ($\kappa\ll 1$) and Type II ($\kappa\gg 1$) regimes.
There is a crossover  $\kappa\equiv \kappa_x(\sim 1)$ that separates these two
extreme regimes. The $\kappa > \kappa_x$ regime shows a
Meissner-like state for  $f<f_{c1}$, with the characteristic field
$f_{c1}<1$,  and it has a  vortex lattice for $f> f_{c1}$.
The vortex lattices resemble the ones that appear in
the ground states of the extreme Type II regime ($\kappa\to\infty$).
For $\kappa <  \kappa_x$ there is Type I-like behavior.
For $f<f_{c0}$,  with $f_{c0} > 1$, the ground state is
a Meissner state. For larger fields $f>f_{c0}$, the field penetrates the array
from the boundaries of the array
in the form of ``vortex collars'' with constant vorticity. The
vortices attract each other to form the collars \cite{nos3,nos3b}.
We stress that the qualitative physical aspects
of the two extreme regimes are independent of the specific inductance model
considered. The difference appears in the form of the long-range
vortex-vortex interaction
which  is generally only relevant for low vortex concentrations.
The quantitative differences arise mainly in the specific values
of the critical fields $f_{c1}$, $f_{c0}$, and the crossover screening
parameter $\kappa_x$ \cite{nos3b}. In the study of the subharmonic giant
Shapiro steps the two physically extreme regimes lead to fractional
giant steps in the IV characteristics with completely different underlying
vortex dynamics.

\section{\bf GIANT SUBHARMONIC SHAPIRO STEPS WITH SCREENING}

\subsection{\bf Coherent vortex oscillating states with local inductance
matrices}

In this subsection we present the results obtained from simulations
of a periodic JJA described by Eq.~(\ref{temp}), with inductance matrices
given by models A or B. Here we concentrate in the zero field case
($f=0$) for the same parameters used in Sec. 3.2. In this case the
dynamical equations were integrated using a second order Runge-Kutta method
with fixed step $\Delta t = 0.02 - 0.1 \tau$, with $\tau$ the smallest
of the characteristic times $\tau_\theta(=1/\nu_\theta$) or
$\tau_\Phi(=1/\nu_\Phi)$.
After discarding the results from the first $100$ periods of the ac current,
time averages were carried out for time intervals of, typically, up to
$500$ periods.

The study of model A with $f=0$ leads to results entirely equivalent to
the one junction case, i.e. the IV characteristics present only
integer  giant Shapiro steps. Significant changes occur, however, when
rational frustrations are applied to the array (see Section 5.3).

The situation for model B with $f=0$ is completely different.
We studied the response of the inductive JJA as a function of
$\kappa_J$, fixing ${\cal M}/\Lambda_0 =0.1$  (i.e. $d=0.4$), and we
found that for all the values of $\kappa_J$ considered {\bf there are}
half-integer GSS.  This appears to happen as soon as  ${\cal M}\neq 0$
(i.e. whenever $E(R)\neq 0$).
Therefore, the presence of the edge magnetic field term is necessary
to induce  subharmonic steps in the IVs.
 From an analysis of a visual animation of the
vortex dynamics, we found that the edge magnetic fields are the basic source
of VAPs that are then responsible for the existence of 1/2-steps in the
IV characteristics.


In Fig.~8 we show the half-integer steps in the IV curves for model B in
the two different regimes of $\kappa_J$. Fig 8(a) has $\kappa_J=2.27$, which
corresponds to a moderate Type II regime. The results in Fig 8(a)
are similar to the ACVS results, obtained when $\kappa=\infty$, in that
they are hysteretic and the 1/2-step width is about the same size in both
cases, for the same parameter values.

Fig.8(b) shows our I vs V results when $\kappa_J=0.44$, i.e. in Type I regime.
Again, there is a 1/2-step but with a hysteresis loop significantly reduced as
compared to the one for $\kappa_J=2.27$. Furthermore, there is clear evidence
of a 2/3-step in the IV, but with larger step width than in the
$\kappa=\infty$ case
and for a smaller lattice size ($40\times 40$).

In Fig.~9 we show the change in the 1/2-step width ($f=0$) as a function of
$1/{\kappa_J}$. The $\kappa_J=\infty$ point, denoted as ACVS, was
generated by including the edge magnetic fields as boundary conditions in
the RSJ JJA equations (See Sec~ 3.5). For finite $\kappa_J$
we can distinguish three qualitatively different regimes in Fig.~9.
(i) the Type II regime, where the 1/2-step width remains essentially constant
and equal to the $\kappa =\infty$ result. (ii) the Type I regime where the
step width first {\it increases} and then (iii) decreases.
The decrease occurs when the characteristic frequency $\nu _{_{\Phi}}<\nu$.
This means that the current distributions are not able to follow the
rapid oscillations due to the external current drive, and thus the coherent
vortex oscillating state is less stable.

We now move to discuss the microscopic coherent vortex patterns responsible
for the 1/2-steps in model B as a function of $\kappa_J$.
Typical examples of the family of oscillating coherent
vortex states found in our studies of model B are shown in  Fig.~10.
In Fig.~10(a) we show results for $\kappa_J =2.27$ ($f=0$).
The oscillating vortex configurations consist of columns of mostly
{\bf isolated} unit charge positive and negative vortices with period $2/\nu$.
As $\kappa_J$ decreases from $\kappa_J =\infty$ to a finite value,
$1<\kappa_J<< \infty$, the ACVS angle for the symmetry axis
changes until it becomes collinear with the direction of the
external current (compare with Fig.~6(a)).
When $\kappa_J <1$, there are two distinct 1/2-step vortex oscillatory
patterns one for $\nu <\nu _{\Phi}$ (ii)  and the other for
$\nu >\nu _{\Phi}$ (iii). In case (ii) the vortex columns are generated
at the sample edges and then move towards the center of the array where they
collide creating and annihilating individual vortices that concentrate about
the center of the array. An instantaneous vortex pattern for
$\kappa_J=0.44$ is shown  in Fig.~10(b). In the regime where
$\nu >\nu _{\Phi}$, the induced currents can not follow the applied current.
Then the vortex columns,  nucleated at the sample edges,
do not have enough time to collide with each other while still oscillating
back and forth with period two. This situation is shown in
Fig.~10(c) for $\kappa_J=0.29$. In  all cases considered,
during the transient regime, VAPs are nucleated
at the boundaries due to the presence of the antisymmetric edge-fields.
But they do not correspond to the nucleation of  {\it commensurate} vortex
states as was suggested in the qualitative explanation of Ref.~\citelow{cinci}.

The difference in the coherent vortex patterns from the
ACVS of Fig.~6(a) to the vortex stripes of Figs.~10(c) is due to the
change in the long-range interaction between vortices, purely logarithmic in
the non inductive ACVS to the exponential decay in model B given in
Eq(~\ref{inteab2}), with the $\kappa_J$-dependent
characteristic decay length from Eq(~\ref{lambdap}). Of significant importance
is the sign of the vortex interactions going from being repulsive for
$\kappa_J=\infty$ to attractive for $\kappa_J <1$
(when compared to the lattice pinning potential \cite{nos3b}).

\subsection{\bf Coherent vortex oscillating states with full inductance matrix}

In this section we discuss our results from including the full-range inductance
matrix in the analysis, within our model C approximation. This is a problem
that has also recently been studied by Phillips {\it et al.} \cite{phillips2},
and we will make comparisons between our results and theirs where appropriate.
We are also  interested in comparing the results discussed in this
section with those of the previous section.
In particular we want to find out how important is the form of the vortex
interaction at large distances, which is where the differences between model
B and C arise, in the formation of coherent oscillating vortex patterns
in zero magnetic field.

In Fig.~11 we show our results including the full inductance matrix of
model C, with $\kappa_p=2$ and for a $24\times 24$ lattice.
The driving ac current and frequency are
the same as in the previous cases.  The numerical integration of the
dynamical equations is done following the  same approach with time step
of $\Delta t=0.1\tau_\theta$. In this case we discarded  the first $20$
periods of the driving current and carried out time averages over $80$
periods. In Fig.~11 we see a clearly developed hysteretic half-integer
Shapiro step. If we compare this IV with the one shown in
Fig.~8(a), for $\kappa_J=2.27$ and only nearest neighbours inductance, we
see that the 1/2-step width is larger in this case.
As mentioned before, the calculations with the full inductance matrix are
more demanding than those of model B and thus we were not able to do
a comprehensive study of the IV steps as a function of $\kappa_p$,
as was done in Fig.~9 for model B.

The important  aspects of Fig.~11 are that: there is clearly a 1/2-step,
there is hysteresis, although its size is intermediate between
the $\kappa_J \gg 1$ and  $\kappa_J <1$ cases, and the non hysteretic part
of the 1/2-step is larger than in model B. These results indicate that
the presence of antisymmetric edge-fields ($E(R)\not=0$),
either modelled by B or C, is responsible for nucleating the
coherent vortex states that lead to 1/2-steps in the IV's.
To further check the relevance of the edge fields, we set $E(R)=0$
in Eq.~(28) and calculated the corresponding IV.  The results
 for model C are shown in Fig.~11 with a dotted line,
using the same $\kappa_p$ and current parameters as before.
We clearly see no evidence for a $1/2$-step. Instead, the IV curve
corresponds to the single junction solution times $N_y$.

There are differences between
models B and C in the specific
vortex patterns formed, as we can see in Fig.~12.
This is not only because the long-range
interactions between vortices are different, but also because for
model C the antisymmetric edge-fields are non-zero throughout the sample
(in which case the strict meaning of ``edge-fields" loses its value.)
Fig.~12(a) shows the rows of isolated vortex and antivortex patterns for
$\kappa_p=20$, as in model B. This coherent vortex state oscillates
with period $2/\nu$,
with vortices and antivortices interchanging their respective positions
after each period $T=1/\nu$. Note that the vortices have a wavy pattern
with a  tendency to form an angle remarkably resembling the one existing
in the ACVS at $\kappa=\infty$.
In Fig.~12(b) we show the vortex pattern for $\kappa_p=2$.
We see that there is one row of vortices and one of antivortices, which
oscillate with period $2/\nu$. After one period $T$, the rows
move back and forth towards the center of the array, but they
do not interchange positions.
In Fig.~12(c) we show the vortex state for
$\kappa_p=0.1$. In this case the large screening makes the
$E(R)$ contribution dominant
(the magnitude of $E$ grows as $1/\kappa_p$).
In this figure we see that there are curved edge-field induced
 rows of vortices (from the left) and antivortices (from the right).
 Only in the center
of the array can we distinguish some isolated vortices and antivortices. This
coherent vortex state also oscillates with period $2/\nu$. After one
period $T=1/\nu$, the edge field induced vortex rows oscillate back and
forth (without changing sign), whereas the isolated vortices and antivortices
close to the center interchange their respective positions. This behavior
is similar to the one seen in Fig.~10(b), but here the edge-field
induced vortex rows occupy most of the sample. However, since the
edge magnetic fields are size dependent, we expect that in real arrays
(which are $1000\times 1000$), the central region with the oscillating
vortices will be bigger.

\subsection{\bf Fractional giant Shapiro steps with screening}

We now proceed to discuss the effects of screening in the stability of the
field induced fractional giant Shapiro steps for $f=p/q$ \cite{nos4}.
In Sec.~2.3 we discussed the $\kappa=\infty$ case where
the driven JJA shows giant fractional voltage steps given in
Eq(~\ref{eq:fgss}). The existence of the FGSS  is due to the collective
oscillation between the different ground state vortex configurations with
frequency  $\nu/q$ \cite{benz,numshap1,numshap2,teorshap}.
The question that naturally arises is what will happen when there is large
screening, i.e. $\kappa<1$, for the ground state now is a Meissner
state \cite{nos3,nos3b}.

In this case, as we show below, we  found that it was enough to use
the diagonal inductance, model A, to simulate the JJA dynamics.
The fields considered were $f=1/3$ and $1/2$. We note that the external field
induces  high vortex concentrations, that in turn
make the effect of antisymmetric edge-fields and differences in the
long-range vortex  interactions less relevant. In Figs. 13(a) and 13(b)
we show the IV results for model A, for a lattice with
$40\times 40$ sites and $f=1/3$.   Fig.~13(a) shows the results
for $\kappa_J=2.27$, while Fig.~13(b) has $\kappa_J=0.44$.
It is clear from these results that in both limits the IV's show giant Shapiro
steps at 1/3 and 1/2 fractions. We note that in the $\kappa_J >1$ regime there
is a small 1/2-step, as was found in the $\kappa =\infty$ case using
free end boundary conditions (Lee and Stroud \cite{numshap1}). For
$\kappa_J <1$ the 1/2-step is much bigger. There are also higher harmonics
seen in Fig.~13(b)  that we have not highlighted in the figure for
clarity. Fig.~13(c) shows the results for model B for the same parameters
as in Fig.~13(b) and with $d=0.4$. A comparison between Figs.~ 13(b) and (c)
shows that qualitatively there are no significant differences.

Fig. 14 shows the widths of the FGSS for $f=1/3$ for steps at
$1/3, 2/3$ and $1/2$, plotted as a function of $1/{\kappa_J}$.
For comparison, in Fig. 14  we also show the results  for
the 1/2-step width with $f=1/2$ as a function of $\kappa_J$.
Again, we distinguish three qualitatively different
$\kappa_J$-regimes in this figure. (i) The Type II regime where the step
width is constant and equal to the $\kappa_J =\infty$ result;
the Type I regime where the step width (ii) increases
initially  and then (iii) decreases. The decrease also occurs when the
characteristic frequency $\nu_{_{\Phi}}<\nu$. As mentioned before
the $FGSS$ in the $\kappa =\infty$ limit were explained in terms of a
coherent oscillation of the ground state vortex lattices formed
when $I^{ext}=0$ \cite{benz,numshap1,numshap2,teorshap}.
When screening is included the vortex lattice configurations get modified.
In  Fig.~15 we show the instantaneous vortex
configurations at the 1/2-step in the $f=1/2$ case, both in the Type II
and Type I cases. In Fig.~15(a) we took $\kappa_J =2.27$, and the vortex state
shown in the figure oscillates with period $2/\nu$.
We notice that there are remanents of the ground states present in the
$\kappa =\infty$ limit but now forming ``clusters"  separated by  ``Bloch"
or ``soliton" surface walls due to the screening currents. The size of
the ``cluster" regions in this particular case is about $\ell =7a$ while the
Bloch wall is
$2a$. As $\kappa_J \rightarrow \kappa _x$, with $\kappa _x\leq 1$
the number of Bloch walls increases, and for $\kappa_J < \kappa_x$
we notice the merging of vortices into the stripes shown in Fig.15(b).
For $\kappa_J=0.44$. This structure is reminiscent of the
intermediate state in a bulk superconductor, although the structure discussed
here is generated dynamically.
The 1/2-step in the
IV-characteristics corresponds to a coherent oscillation of
this stripped vortex distribution that oscillates between $+1$ and
zero vorticities with period $2/\nu$.
In the Type I regime with $I^{ext}=0$, the equilibrium magnetic field
distributions   correspond to the Meissner state \cite{nos3,nos3b}.
Starting from this state and
turning on the external $I^{ext}(t)$ current,
the induced Lorentz force   ``pulls" the flux
into the sample and, after a transient, the vortex distributions
settles into the final oscillating stripped configuration.
This is shown in Fig.~16 for $f=1/3$, within the $1/3$-step.

In both Type I and Type II limits, the oscillating vortex states are
non-equilibrium stationary states.
In the Type II regime the oscillating vortex lattice, even when similar to the
ground state of the $\kappa_J=\infty$ case, is different because of the
screening induced defects. The vortex state
without external driving, $I^{ext}=0$, forms a lattice but with a
density of vortices smaller than the external field density $f=p/q$,
because there is now a finite penetration depth \cite{nos3,nos3b}. On the other
hand, the oscillating vortex state at the fractional steps, has a
vortex density exactly equal to $f=p/q$, so that it can
oscillate with frequency $\nu/q$. Only in the $\kappa=\infty$ limit
both the oscillating vortex state and the ground state coincide.
In the Type I regime, as it was mentioned before, the difference
is more explicit. Whereas the ground state is the Meissner state, without
vortices, the oscillating vortex state is the ``intermediate'' state
shown in Fig.~15(b) and Fig.~16(d).

Although we did not carry out calculations for $f=p/q$ with model C we do
not expect significant qualitative changes as compared to the picture
presented above,
since for high vortex concentrations we have not found significant
differences between models A and C in the static case \cite{nos3b}.

\subsection{\bf Magnetic field dependence of half-integer Shapiro steps}

In the experiments by H. C. Lee {\it et al.} \cite{cinci} it was found that
the 1/2-steps exist not only for zero field, but also for any finite
value of $f$. As we discussed above, we have found that as long as we include
the antisymmetric edge-fields, represented by a $E(R)\not=0$,
there are 1/2-steps in the IV's for all the values of $f$ considered.
In Fig.~17 we show the step width of the 1/2-integer step as a function
of $f$ both for a Type II ($\kappa_J=2.27$)
and Type I cases ($\kappa_J=0.44$). In this case we simulated model B
with $d=0.4$. In the Type II case the curve shows more structure than in
the  Type I regime. Also, in the Type II regime
the  1/2-step width close to $f=0$ tends to go to zero for
small fields. Moreover, when the ACVS is induced by disorder
($\kappa=\infty$) we  found that the 1/2-step disappears for fields
smaller than $f=0.05$.  What is qualitatively important
is that in both regimes the curves have maxima at $f=0$ and $f=0.5$, as
observed in the experiments \cite{cinci}. The relative height of the two
maxima depends on the frequency and amplitude of the ac current.

Although all the results presented above are $T=0$ results, we have verified
that they are stable at sufficiently  low temperatures.

\section{\bf CONCLUDING REMARKS}

The central theme of this review was the search for a microscopic
understanding of the different types of steps that can be
found in the IV characteristics of driven proximity  effect Josephson
junction arrays (JJA). The motivation for this work started with the
experimental findings of giant integer ($f=0$) and fractional ($f=p/q$)
Shapiro steps \cite{gshap1,benz} (GSS), that were theoretically explained
in terms of
an extreme Type II RSJ model \cite{numshap1,numshap2,teorshap}.
The dynamical equations of motion describing the JJA are equivalent to
a large set of overdamped nonlinear coupled equations driven by an external
time-dependent current. However, in the $f=0$ case the
array behaves as a set of single junctions connected in
series, and thus the GSS
are simply a manifestation of the Shapiro steps of each junction.
Note that if we were discussing the underdamped limit of a single junction
extra subharmonic states would appear.
When $f=p/q$ the steps seen in the IV's are due to a coherent oscillation
of the ground state vortex lattice \cite{benz,numshap1}.

Experimentally, there are fractional giant Shapiro steps in zero field
that do not conform to the picture given
above\cite{benz,cinci,garland,garland2}.
In order to shed light into this problem we decided to study the effects
of disorder in a driven JJA modeled by the RSJ model
in zero field and $\kappa=\infty$ \cite{nos,nos2,nos2b}. We discovered that
as soon as we
break translational invariance the corresponding IV's show half-integer
GSS. When looking in more detail at the corresponding microscopic vortex
dynamics, we found a stationary coherent oscillatory vortex pattern
with completely
unexpected properties. We termed this novel state {\it axisymmetric coherent
vortex state} (ACVS), which characterizes its geometric
nature. We have carried out a comprehensive
analysis of the stability properties of the ACVS against variations of several
physical parameters. We have concluded that the ACVS is in fact a very stable
global attractor of the nonlinear dynamics, that can be generated in
several different
ways. The only essential element needed to nucleate the ACVS, however,
is to have a mechanism that produces a vortex antivortex pair (VAP) in the
initial conditions.

Unfortunately, we still do not have a complete theoretical understanding of the
ACVS, in particular, the ``radiation" and annihilation mechanisms that
are essential for its formation and stability.
The VAP radiation phenomenon, due to the presence of the ac current,
does not appear to have been studied before. Furthermore, we would like to
understand why the ACVS forms in the very specific patterns shown in the
figures. Note that the ACVS is a true nonequilibrium state that disappears
as soon as we turn off both the dc and/or ac currents.
This property seems to imply that the ACVS can not be explained from a
minimum energy principle. This is a fundamental difference between the ACVS
and the fractional GSS observed with $f=p/q$.\cite{numshap1,numshap2,teorshap}

Most of the experiments that show GSS were made in proximity effect
arrays, with large screening currents. We began the study of self-field
effects in JJA by considering local screening models for the inductance
matrix\cite{nos4}. There we found that there are new fractional GSS triggered
by the current induced  antisymmetric magnetic edge-fields. We found that
there is a robust 1/2-step in the IV's even when $f=0$, and similar to what
was seen in the experiments.

Recently, Phillips {\it et al.} \cite{phillips} and Reinel
{\it et al.} \cite{majhofer2} have extended the static analysis including
the full inductance matrix, and then
Phillips {\it et al.} \cite{phillips2} extended our dynamic study including
the full inductance matrix. The inductance matrix model used by
Phillips {\it et al.} takes fully into account the specific geometry of
typical SIS arrays.  The local geometry of these arrays is quantitatively
different from the SNS junctions used in the
study of GSS \cite{benz,cinci,garland,garland2}. However, their results
must be semi-quantitatively correct, in particular at long distances.

In this review we have presented our own full inductance matrix
results within the filamentary approximation.  We have discussed
the main differences between  local screening models (A and B)
and the full inductance model (C).
Models  A and B differ in that model A does not include
``edge-fields'' ($E(R)=0$) while model B does ($E(R)\not=0$).
Nonetheless, as shown in Section 4.3, both models have, within the
linearized approximation of the equations of motion, the same long distance
vortex-vortex (V-V) interaction potential. On the other hand,
models B and C have different long distance
V-V interaction potential. In model B the  V-V interaction
decays exponentially with distance, while in model C
it decays algebraically.
Nevertheless, both models include the ``edge-fields''
 (quantitatively  the range of $E(R)$ depends on the
range of $L(R,R')$.)

 From our comparisons of the three inductance matrix models, the following
picture emerges:

(i) The existence of the subharmonic GSS at $f=0$ is due to the
presence of current induced antisymmetric magnetic fields,
which break the translational
symmetry. In fact, when we artificially  imposed $E(R)=0$ in model
C, we did not obtain a subharmonic response, despite of  the fact
that the long range V-V interactions were properly taken into account.
Instead, we get the single junction type  IV (see Fig. 11.)

(ii) The specific structure of the underlying coherent oscillating vortex
states and their corresponding
subharmonic step widths, depend directly on the long
range V-V interactions (as also found by Phillips {\it et al.}.)
We found that as the screening parameter $\kappa$ decreases,
there is a whole family of coherent oscillating
vortex states where  the vortex rows tend to line up with the
external current, for a given inductance model.
For $\kappa$ larger than the crossover $\kappa_x(\sim 1)$,
the V-V interaction is repulsive while for
$\kappa\leq\kappa_x$ it is attractive. This explains why the
vortex patterns in the former case contain isolated vortices while in the
latter the vortices form stripes of constant vorticity.
Also, the  geometric
vortex patterns depend on the model used to describe the inductance matrix
for a given $\kappa$,
which is connected to the difference in the nature of the V-V interaction.

In all the cases studied (either with $f=0$
or $f=p/q$) the coherent vortex states are  a fundamentally
non-equilibrium consequence of the dynamics,
with the only exception being the particular case
of $\kappa=\infty$ and $f=p/q$. We believe that these
non-equilibrium coherent vortex states, due to the breaking of translational
invariance, are the underlying mechanism
behind all the fractional giant Shapiro steps observed in {\it
two-dimensional} SNS JJA.
At $f=p/q$ this translational symmetry
is broken by the applied magnetic field, that generates higher order
periodicity  thus leading to {\sl fractional} GSS.
At $f=0$ the translational symmetry can be broken either by defects or by
the current induced antisymmetric magnetic fields. They tend
to nucleate vortex-antivortex pairs that produce
ACVS-like vortex patterns responsible for the {\sl subharmonic} GSS.
Fractional steps can also be generated  in lower dimensional systems by
breaking translational invariance (e.g. ladder arrays and single plaquettes
\cite{ladder,korea1,korea2,octavio}).

The subharmonic GSS studied here correspond to fully overdamped
JJA. It is known that underdamped single Josephson junctions
can have subharmonic Shapiro steps and even chaos \cite{chaos}.
In fact, some chaotic phenomena has been recently studied in
ac driven underdamped JJA \cite{chaos2}. It would be
interesting also to know the effect of a finite capacitance (but
not large enough to induce chaos) in these coherent vortex states,
together with a finite vortex mass and the effects of the non-linear
vortex viscosity \cite{holand3}.

We also want to mention other possibly related systems,
like charge density waves \cite{cdw} and the Frenkel-Kontorova model
\cite{frenkel}, where subharmonic Shapiro
steps have been found as a consequence of the collective nonlinear
dynamics of these systems.

How can the presence of a specific coherent oscillatory vortex state
be verified experimentally?  There are several possibilities but
we will only mention the ones that are closest to the JJA system.
Recently, many different techniques have been developed for doing
spatially resolved measurements in JJA \cite{pannetier,urbana,hallen,susana}
(save for Tonomura's {\it al.} approach\cite{tonomura}.)
Most of these techniques, however, are dedicated to the study of
{\it static} flux configurations \cite{pannetier,urbana,hallen}.
Lachenmann {\it et al} \cite{susana} have been able to study {\it dynamic}
states in JJA by the using the low temperature scanning electron
microscopy (LTSEM) technique. These are basically
measurements of the local distribution of dissipation produced by the
local heating of the LTSEM tip, that in turn gives the  average
junction voltage in the arrays.
Since a moving vortex generates a voltage as it crosses a junction,
this type of measurement indicates the location of the vortices
during their oscillations (note that since
this process is independent of the sign of the vorticity, it will not average
to zero after many ACVS ac current periods.)
In fact, in Fig.~18, we show a simulation of the average voltage
along the current direction for the ACVS at a finite temperature. There
we see where the vortices were located most of the time regardless of their
sign. In particular, the characteristic angle of the ACVS is clear.
Therefore, an LTSEM based measurement may be able to provide evidence
for the formation of an ACVS.

In conclusion, we have shown
that giant subharmonic steps in the IV characteristics
can be generated under many different circumstances at zero and
non-zero external magnetic field,
but for which the underlying microscopic vortex dynamics can
be fundamentally different. In all the cases considered, at least for two
dimensional SNS arrays, the steps appear to always entail a microscopic
coherent  oscillating vortex state.

We do not as yet have a more theoretical understanding of all the  nonlinear
mechanisms responsible for the formation of all these coherent vortex states
and further studies are needed to shed light on these questions.

\nonumsection{Acknowledgments}

We thank  A.\ Karma and C.\ Wiecko for their collaboration in the
initial stages of our ACVS studies. Fruitful discussions with  H. van
der Zant, R.\ Markiewicz, J. E. van
Himbergen, M.\ Rzchowski, D.\ Stroud, A.\ Majhofer, T. Hagenaars,
P. Tiesinga, and L.L. Sohn are gratefully acknowledged. One of us (JVJ) thanks
the Instituut voor Theoretische Fysica,
Universiteit Utrecht, The Netherlands, for its kind hospitality during
his 93-94 sabbatical leave, specially J. E. van Himbergen.
We thank  the $NSF$ grant DMR-9211339, the  Pittsburgh supercomputing center
under grant PHY88081P, the Donors of the Petroleum Research Fund,
under grant ACS-PRF\#22036-AC6, and the ICTP, Trieste,
for partial support of this research.
D.D. was also supported by a fellowship from CONICET, Argentina.

\nonumsection{References}

\newpage

\section{Figure Captions}

\vspace{0.5cm}

\noindent
{\bf Figure 1:} Schematic representation of a Josephson junction square array
driven by an external current $I$. The sites denote superconducting islands
and the crosses the junctions themselves.

\vspace{1cm}
\noindent
{\bf Figure 2:} Vortex configurations, for a 40$\times 40$ array with one
defect, produced by a dc current with $f=10$, $\delta_x = 0.1$, $\delta_y=0$.
The white and black squares denote $-1$ and $+1$ phase
vorticity per plaquette (see Eq.~ (11)), respectively. No squares here
means zero vorticity.  For  currents (a) $I_{dc} =
0.8I_0 < I_c =0.84I_0$; (b) $I_{dc}=0.87I_0 > I_c$.

\vspace{1cm}
\noindent
{\bf Figure 3:} IV characteristics for a $40\times 40$ square array with
$I_{ac}= I_0$, $\nu=0.1\nu_0$ and pbc. (a) Ordered array; (b) array with
one defect in the center of the lattice and $\delta_x=0.1$,
$\delta_y=0$ and $f=2$
(for clarity the IV curve is shifted by one unit); (c) blowup of (b) for an
$80\times 80$ array showing hysteretic behavior and a clear 2/3-step.

\vspace{1cm}
\noindent
{\bf Figure 4:} Step width for of the 1/2-step for arrays with one
defect given by the same parameters as in Fig.~3(b).
(a) As a function of the frequency, ($I_{ac} = I_0$);
and (b) as a function of the ac current, ($\nu = 0.1 \, \nu_0$).

\vspace{1cm}
\noindent
{\bf Figure 5:} Number of vortices as a function of time for an array with
one defect and $I_{dc} = 0.586I_0$ (within the 1/2-integer
step). The lattice size, frequency and ac current are the same as in
Fig.~3 (b). (a) Absolute number of vortices $N_a(t)$; (b) total
number of vortices $N_T(t)$. See text for definitions of $t_d$,
$t_{ACVS}$ and $t_{off}$. Here time is measured in units of
$1/\nu_0$.

\vspace{1cm}
\noindent
{\bf Figure 6:} Axisymmetric coherent vortex states (ACVS). (a-b) Stationary
oscillating vortex configuration for an ACVS in the 1/2-step,
for the same parameter values as in Fig.~3(b)
with $I_{dc}=0.586I_0$.
(c) ACVS configuration in the  2/3-step with $I_{dc}=0.622I_0$
in an $80\times 80$ lattice. We use the same conventions as
in Fig.~2.

\vspace{1cm}
\noindent
{\bf Figure 7:} Vortex distributions showing an ACVS at a 1/2-integer
step generated by antisymmetric edge-fields in a $64\times 64$ lattice with
$\gamma = 1$, $I_{ac}=I_0$, $\nu = 0.1\nu_0$ and
$I_{dc}=0.57I_0$. The notation is the same as in Fig. 2.

\vspace{1cm}
\noindent
{\bf Figure 8:} IV-characteristics  for arrays with only nearest-neighbours
inductance (model B): Here $f=0$, $I_{a.c.}=I_0$,
$\nu = 0.1 \nu_{\theta}$, and ${\cal M}=0.1\Lambda_0$ ($d=0.4$),
for  $40\times 40$ lattice size. (a) Type II array
with $\kappa_J = 2.27$; (b) Type I array with $\kappa_J = 0.44$.

\vspace{1cm}
\noindent
{\bf Figure 9:} IV-step widths ${\Delta I}/I_0$ vs $1/\kappa_J$, for
half-integer steps in zero field.
$I_{a.c.}=I_{0}, \nu = 0.1\nu_{\theta},{\cal  M}=0.1\Lambda_0$, and lattice
size  $40\times 40$.

\vspace{1cm}
\noindent
{\bf Figure 10:} Instantaneous vortex configurations for 1/2-steps with $f=0$
(model B).  Same ${\cal M}$ and ac driving currents as in Fig.~9, and
$I_{dc}=0.61I_0$.
(a) $\kappa_J = 2.27$; (b) $\kappa_J=0.44$; and (c) $\kappa_J = 0.29$.
 We use the same conventions as in Fig.~2 to denote vorticity.

\vspace{1cm}
\noindent
{\bf Figure 11:} IV-characteristics of
 an array with full inductance matrix (model C): $\kappa_p=2$, lattice
size $24\times 24$,  $f=0$, $I_{a.c.}=I_0$, and
$\nu = 0.1 \nu_{\theta}$. The dotted line corresponds to a simulation
with the same model C inductance, same $\kappa_p$ and applied currents,
but without edge fields  (i.e. imposing $E(R)=0$ in Eq.(24)).

\vspace{1cm}
\noindent
{\bf Figure 12:} Instantaneous vortex configurations for 1/2-steps at $f=0$,
 $I_{ac}=I_0$, $\nu=0.1\nu_\theta$, and $I_{dc}=0.59I_0$, for model C.
(a) $\kappa_p = 20$; (b) $\kappa_p=2$, (c) $\kappa_p=0.1$.
 The same conventions to indicate vorticity as in Fig.~2 is used.

\vspace{1cm}
\noindent
{\bf Figure 13:} IV-characteristics for model A with $f=1/3$, $I_{a.c.}=I_0$,
and $\nu =0.1 \nu_{\theta}$, for lattice size $40\times 40$. (a) Type II array
with $\kappa_J = 2.27$; (b) Type I array with $\kappa_J = 0.44$; (c) The
same as (b) including edge-fields (model B)
with ${\cal M}/{\Lambda_0} =0.1$  ($d=0.4$). The giant Shapiro steps
at $n/3$ and $n/2$ are marked. Curves start about the same place and are
displaced from each other for clarity.

\vspace{1cm}
\noindent
{\bf Figure 14:} IV-step widths, ${\Delta I}/I_c$ vs $1/\kappa_J$,
for $I_{a.c.}=I_{0}, \nu = 0.1\nu_{\theta}$, diagonal inductance
approximation (model A), and lattice size $40\times 40$.
Upper curve corresponds to 1/2-step for $f=1/2$, while the three lower
curves to 2/3-step, 1/3-step and 1/2-step with $f=1/3$.

\vspace{1cm}
\noindent
{\bf Figure 15:} Instantaneous vortex configurations for 1/2-steps
with $f=1/2$, in the diagonal inductance approximation (model A). $I_{ac}=I_0$,
$\nu=0.1\nu_\theta$. (a) $\kappa_J =2.27$ (Type II), with $I_{dc}=0.4I_0$.
(b) $\kappa_J = 0.44$ (Type I), with $I_{dc}=0.6 I_0$.
White and black squares indicate $0$ and $+1$ phase vorticity per plaquette,
respectively.

\vspace{1cm}
\noindent
{\bf Figure 16:} Transient time evolution for the vortex configurations
at a 1/3-step for $f=1/3$, in model A. Here,
$I_{dc}=0.45I_0$, $I_{ac}=I_0$, and $\nu=0.1\nu_\theta$. The current is
applied along the vertical direction.
(a) Initial Meissner configuration (i.e. the ground state for $I_{ext}=0$),
 (b) The first vortex row has penetrated the array. (c) Advancing front of
 vortex rows. (d) Final oscillating 1/3-period state of vortex rows.
White and black squares indicate $0$ and $+1$ phase vorticity per plaquette,
respectively.

\vspace{1cm}
\noindent
{\bf Figure 17:} Current width of the 1/2-steps
as a function of $f$ (model B)
 for: (a) $\kappa_J=2.27$ and (b) $\kappa_J = 0.44$.
Same ${\cal M}$ and driving currents as in Fig.~9.

\vspace{1cm}
\noindent
{\bf Figure 18:} Time average of the voltage drop in each junction along
the current direction. Contour plot for the case with $I_{dc}=0.57I_0$,
$I_{ac}=I_0$, $\nu=0.1\nu_\theta$ (i.e. in the half-integer step), for
a $40\times 40$ lattice with $\kappa=\infty$ and finite temperature
$k_BT=0.01E_J$.


\begin{thebibliography}{23}
\bibitem{nota} Note that in the experiments of
Ref.~\citelow{benz,cinci,garland,garland2} frequencies have been
normalized by $\nu_c=\nu_0/2\pi$, therefore $\nu_{max}\approx 1.38\nu_c$
in their notation.
\bibitem{nota2}
It is worth to mention that Hebboul and Garland \cite{garland2}
have observed  $1/2$-steps only above a critical $I_{ac}$
in their experiments. However, as we discuss later on in this article, it
is more likely that in their case the subharmonic steps are due
to the presence of antisymmetric current induced magnetic fields, instead
of defects.
\end{thebibliography}
\end{document}